\documentclass[reprint,
superscriptaddress,
amsmath,
amssymb,
aps,
twocols
]{revtex4-1} 
\usepackage{verbatim}
\usepackage{graphicx}
\usepackage{dcolumn}
\usepackage{bm}
\usepackage{mathtools}                                
           
\usepackage{float}                                    
      
\newcommand\asmap{\ensuremath{\Sigma}}
\usepackage{graphicx}                                 
\usepackage{caption}                                 
\usepackage{subcaption}                               
\usepackage{hyperref}        
\usepackage{multirow}                         
\usepackage[usenames, dvipsnames]{color}
\usepackage{color}
\usepackage{xr}

\newcolumntype{P}[1]{>{\centering\arraybackslash}p{#1}}

\begin{document} 

\title{An information theory--based approach for optimal model reduction of biomolecules} 

\author{Marco Giulini}
\affiliation{Physics Department, University of Trento, via Sommarive, 14 I-38123 Trento, Italy}
\affiliation{INFN-TIFPA, Trento Institute for Fundamental Physics and Applications, I-38123 Trento, Italy}
\author{Roberto Menichetti}
\affiliation{Physics Department, University of Trento, via Sommarive, 14 I-38123 Trento, Italy}
\affiliation{INFN-TIFPA, Trento Institute for Fundamental Physics and Applications, I-38123 Trento, Italy}
\author{M. Scott Shell}
\affiliation{Department of Chemical Engineering, University of California Santa Barbara, Santa Barbara, California 93106, USA}
\author{Raffaello Potestio}
 \email{raffaello.potestio@unitn.it}
\affiliation{Physics Department, University of Trento, via Sommarive, 14 I-38123 Trento, Italy}
\affiliation{INFN-TIFPA, Trento Institute for Fundamental Physics and Applications, I-38123 Trento, Italy}

\date{\today}

\begin{abstract}
In the theoretical modelling of a physical system a crucial step consists in the identification of those degrees of freedom that enable a synthetic, yet informative representation of it. While in some cases this selection can be carried out on the basis of intuition and experience, a straightforward discrimination of the important features from the negligible ones is difficult for many complex systems, most notably heteropolymers and large biomolecules.
We here present a thermodynamics-based theoretical framework to gauge the effectiveness of a given simplified representation by measuring its information content. We employ this method to identify those reduced descriptions of proteins, in terms of a subset of their atoms, that retain the largest amount of information from the original model; we show that these highly informative representations share common features that are intrinsically related to the biological properties of the proteins under examination, thereby establishing a bridge between protein structure, energetics, and function.
\end{abstract}

\maketitle

\section{Introduction}

The quantitative investigation of a physical system relies on the formulation of a {\it model} of it, that is, an abstract representation of its constituents and the interactions among them in terms of mathematical constructs. In the realisation of the simplest model that entails all the relevant features of the system under investigation, one of the most crucial aspects is the determination of its level of detail. The latter can vary depending on the properties and processes of interest: the quantum mechanical nature of matter is explicitly incorporated in {\it ab initio} methods \cite{CarParrPRL1985}, while effective classical interactions are commonly employed in the the all-atom force fields used in all-atom (AA) molecular dynamics (MD) simulations \cite{md_general_method, md_sim_biomol}. Representations of a molecular system whose resolution level is lower than the atomistic one are commonly dubbed {\it coarse-grained} (CG) models \cite{Takada2012,noid_persp,Saunders2013,Potestio2014,d2015coarse}: in this case, the fundamental degrees of freedom, or effective interaction centroids, are representatives of groups of atoms, and the interactions among these CG sites are parametrised so as to reproduce equilibrium properties of the reference system.

An important distinction should be made between {\it reproducing} a given property, and {\it describing} it. For example, it is evident that the explicit incorporation of the electronic degrees of freedom in the model of a molecule is necessary to reproduce, with qualitative and quantitative accuracy, its vibrational spectrum; on the other hand, the latter can be measured and described from the knowledge of the nuclear coordinates alone, i.e. from the inspection of a {\it subset} of the system's degrees of freedom. This is a general feature, in that the {\it understanding} of a complex system's properties and behaviour can typically be achieved in terms of a reduced set of variables: statistical mechanics provides some of the most recognisable examples of this, such as the description of systems composed of an Avogadro number of atoms or molecules in terms of a handful of thermodynamical parameters.

In computer-aided studies, and particularly in the fields of computational biophysics and biochemistry, recent technological advancements--most notably massive parallelisation \cite{foldathome}, GPU computing \cite{gpugrid}, and tailor-made machines such as ANTON \cite{Shaw2009}--have extended the range of applicability of atomistic simulations to molecular complexes composed of millions of atoms \cite{Freddolino2006,Bock2013,SINGHAROY20191098}; even in absence of such impressive resources, it is now common practice to perform microseconds-long simulations of relatively large systems, up to hundred thousands atoms. However, a process of filtering, dimensionality reduction, or feature selection is anyhow required in order to distill the physically and biologically relevant information from the immense amount of data it is buried in.

The problem is thus to identify the most synthetic picture of the system that entails all and only its important properties: an optimal balance is sought between parsimony and informativeness. This objective can be pursued making use of the language and techniques of bottom-up coarse-grained modelling \cite{noid_persp, noid_mapping}; in this context, in fact, one defines a {\it mapping operator} ${\bf M}$ that performs a transformation from a high-resolution configuration ${\bf r}_i$, $i = 1,...,n$ of the system described in large detail to a simpler, {\it coarser} configuration ${\bf R}_I$, $I = 1,...,N < n$ at lower resolution:

\begin{equation}
\label{eq:mapping_firstdef}
    {\bf M}_{I}({\bf r}) = {\bf R}_{I} = \sum_{i=1}^{n} c_{Ii}  {\bf r}_i,
\end{equation}
where $n$ and $N$ are the number of atoms in the system and the number of CG sites chosen, respectively.
The linear coefficients $c_{Ii}$ in Eq.~\ref{eq:mapping_firstdef} are constant, positive and subject to the normalisation condition $\sum_{i} c_{Ii} = 1$ to preserve translational invariance. Furthermore, coefficients are generally taken to be {\it specific} to each site \cite{noid_mapping}, that is, an atom $i$ taking part to the definition of CG site $I$ will not be involved in the construction of another site $J$ ($c_{Ji} = 0 \ \forall \ J \ne I$).

Once the \emph{mapping} ${\bf M}$ is chosen, effective interactions among CG sites must be determined. In this respect, several methodologies have been devised in the past decades to parametrise such CG potentials \cite{Takada2012,noid_persp,Saunders2013,Potestio2014,d2015coarse}. Here, however, we do not tackle this issue, but rather focus on the consequences of the simplification of the system description even if the underlying physics is the same, i.e. configurations are sampled with the reference, all-atom probability. In other words, we focus purely on the effect of projecting the all-atom conformational ensemble onto a coarse-grained configurational space using the mapping as a filter.

Inevitably, in fact, a CG representation loses information about the high-resolution reference \cite{noid_persp, jinJCTC2019}, and the amount of information lost depends only on the number and selection of the retained degrees of freedom. In coarse-grained modelling, the mapping is commonly chosen based on general and intuitive criteria: for example, it is rather natural to represent a protein in terms of one single centroid per amino acid (usually the choice falls on the $\alpha$ carbon of the backbone) \cite{kmiecik2016coarse}. However, it is by no means assured that a given representation that is natural and intuitive to the human eye is also the one that allows the CG model to retain the largest amount of information about the original, higher-resolution system \cite{potestio_jctc, KhotJCP2019}. A quantitative criterion to assess how much detail is lost upon structural coarsening is thus needed in order to perform a sensible choice.

In the past few years, various methods have been developed that target the problem of the automated construction of a simplified protein's representation at a resolution level lower than atomistic. In a pioneering work Gohlke and Thorpe proposed to partition a protein in few, size-wise diverse blocks, distributing the amino acids among the different domains so as to minimise the degree of internal flexibility of the latter \cite{Gohlke2006}. This picture of a protein subdivided in {\it quasi-rigid domains}, which has been further developed by several other authors \cite{ZHANG_BJ_2008,ZHANG_BJ_2009,Potestio2009,aleksiev2009,ZHANG_JCTC_2010,Sinitskiy2012,Polles2013}, is founded on the notion of a simplified model where groups of atoms are not assigned to coarse-grained sites according to their chemistry (e.g., one residue - one site), but rather based on the local properties of the specific molecule under examination. These partitioning methods, however, only employ structural information, in that they aim at minimising each block's internal strain, while the energetics of the system is neglected.

Alternative approaches systematically reduce the number of atoms in a system's representation by grouping them according to graph-theoretical procedures, e.g., mapping the static structure on a graph and hierarchically decimating it by clustering together the ``leaves'' \cite{depabloJCTC2019}, or lumping residues in effective sites based on a spectral analysis of the graph Laplacian \cite{PONZONI20151516}.

More recently, it was proposed to retain only those atoms that guarantee the set of new interactions to be as close as possible to the old ones \cite{koehlJCTC2017,potestio_jctc}. These methods, though, are based on linearised elastic network models \cite{Tirion1996,Bahar1997,Hinsen1998,Atilgan2001,Delarue2002,Micheletti2004} that have the remarkable advantage of being exactly solvable, but cannot be taken as significant representations of the system's highly nonlinear interactions.

It follows that all these pioneering approaches rely either on purely geometrical/topological information obtained from a single, static structure; or on an ensemble of structures, neglecting energetics and thermodynamics; or on extremely simplified representations of both structure and interactions, that do not guarantee general applicability to systems of great complexity.

Here we tackle the issue of the automated, unsupervised construction of the most informative simplified representation of biological macromolecules in purely statistical mechanical terms, that is, in the language that is most naturally employed to investigate such systems. Specifically, we search for the mapping operator that, for a given number of atoms retained from the original all-atom model, provides a description whose information content is as close as possible to the reference. In this context, then, the term ``coarse-grained representation'' should not be interpreted as a system with effective interactions whose scope is to reproduce a certain property, phenomenon, or behaviour; rather, the representations we discuss here are simpler {\it pictures} of the reference system evolving according to the reference microscopic Hamiltonian, but {\it looked at} in terms of fewer degrees of freedom. Our objective is thus the identification of the most informative simplified picture among those possible.

To this end, we make use of the concept of {\it mapping entropy}, $S_{map}$ \cite{Shell2008,rudzinski_2011,Shell2012,foley2015impact}, a quantity that measures the quality of a CG representation in terms of the ``distance'' between probability distributions---the Boltzmann distribution of the reference, all-atom system, and the equivalent distribution when the AA probabilities are projected into the CG coordinate space. The mapping entropy is ignorant of the parametrisation of the effective interactions of the simplified model: $S_{map}$ effectively compares the reference system, described through all its degrees of freedom, to the same system in which configurations are viewed through ``coarse-graining lenses''. The difference between these two representations only lies in the resolution, not in the microscopic physics.

In the following we illustrate a computationally effective protocol that enables the approximate calculation of the mapping entropy. In analogy with the work of Ref.~\cite{foley2015impact}, we employ this novel scheme to identify those representations of the reference molecular system that feature the lowest mapping entropy---that is, allowing for the smallest amount of information loss upon resolution reduction. The method is applied to three proteins of substantially different size, conformational variability, and biological activity. We show that the choice of retained degrees of freedom, guided by the objective of preserving the largest amount of information while reducing the complexity of the system, highlights biologically meaningful and {\it a priori} unknown structural features of the proteins under examination, whose identification would otherwise require computationally more intensive calculations or even wet lab experiments.

\section{Results}

In this section we report the main findings of our work. Specifically, (\emph{i}) we outline the theoretical and computational framework that constitutes the basis for the calculation of the mapping entropy; (\emph{ii}) we illustrate the biological systems on which we apply the method; and (\emph{iii}) we describe the results of the mapping entropy minimisation for these systems and the properties of the associated mappings.

\subsection{Theory}

The concept of mapping entropy as a measure of the loss of information inherently generated by performing a CG'ing procedure on a system was first introduced by one of us in the framework of the relative entropy method~\cite{Shell2008}, and subsequently expanded in Refs.~\cite{rudzinski_2011,Shell2012,foley2015impact}. For the sake of brevity, we here omit the formal derivation connecting relative entropy and mapping entropy as well as a discussion of the former. A brief summary of the relevant theoretical results presented in Refs.~\cite{Shell2008,rudzinski_2011,Shell2012,foley2015impact} is provided in Appendix~\ref{app:reltomap}.

In the following we restrict our analysis to the case of decimation mappings ${\bf M}$, in which a subset of $N<n$ atoms of the original system is retained while the remaining ones are integrated out, so that
\begin{eqnarray}\label{eq:decimation}
&&{\bf M}_I({\bf r}) = \sigma_i {\bf r}_i,\ \sigma_i =  1\ \mbox{for one $I$, 0 otherwise},\\ \nonumber
&&\sum_{i = 1}^n \sigma_i = N.
\end{eqnarray}

In this case, the mapping entropy $S_{map}$ reads (see Appendix~\ref{app:reltomap}) \cite{rudzinski_2011}
\begin{eqnarray}
\label{eq:smap_main}
S_{map} &=&  k_B\times D_{KL}(p_{r}({\bf r})||\bar{p}_r({\bf r})) \nonumber \\
&=& k_B\int d{\bf r}\ p_r({\bf r}) \ln \left[ \frac{p_r({\bf r})}{\bar{p}_r({\bf r})} \right],
\end{eqnarray}
that is, a Kullback-Leibler (KL) divergence $D_{KL}$ \cite{kullback1951information} between the probability distribution $p_r({\bf r})$ of the high-resolution system and the distribution obtained by observing the latter through ``coarse-graining glasses'', $\bar{p}_r({\bf r})$. Following the notation of Ref.~\cite{rudzinski_2011}, $\bar{p}_r({\bf r})$ is defined as
\begin{equation}
\label{eq:omega}
\bar{p}_r({\bf r}) = {p_R({\bf M}({\bf r}))}/{\Omega_1({\bf M}({\bf r}))},
\end{equation}
where $p_R({\bf R})$ is the probability of the CG macrostate ${\bf R}$, given by
\begin{eqnarray}
\label{eq:pmacro1}
 p_R({\bf R})&=&\int d{\bf r}\ p_{r}({\bf r})\delta({\bf M}({\bf r}) - {\bf R}) \nonumber \\
 &=&\frac{1}{Z}\int d{\bf r}\ e^{-\beta u({\bf r})}\delta({\bf M}({\bf r}) - {\bf R}), \nonumber \\
 Z &=&\int d{\bf r}\ e^{-\beta u({\bf r})},
\end{eqnarray}
while $\Omega_1({\bf R})$ is defined as
\begin{equation}
\label{eq:omega1}
\Omega_1({\bf R}) =   \int d{\bf r}\  \delta({\bf M}({\bf r}) - {\bf R}),
\end{equation}
which is the degeneracy of the macrostate---how many microstates map onto the CG configuration ${\bf R}$. In Eq. \ref{eq:pmacro1} $\beta=1/k_BT$, $u({\bf r})$ is the microscopic potential energy of the system, and $Z$ its canonical partition function.

The calculation of $S_{map}$ in Eq.~\ref{eq:smap_main} thus amounts at determining the distance (in the KL sense) between two, although both microscopic, conceptually very different distributions. In contrast to $p_r({\bf r})$, Eq.~\ref{eq:omega} displays that $\bar{p}_r({\bf r})$ associates, to all configurations that map onto the same CG macrostate {\bf R}, the same probability; the latter is given by the average of the original probabilities of these microstates. Importantly, $\bar{p}_r({\bf r})$ represents the high-resolution description of the system that would be accessible \emph{only starting} from its low-resolution one---i.e., $p_R({\bf R})$. Grouping together configurations into a CG macrostate has the effect of flattening the detail of their original probabilistic weights. An attempt to revert the CG'ing procedure and restore an atomistic resolution by reintroducing the mapping operator ${\bf M}$ in $p_R({\bf R})$ can only result in microscopic configurations that are uniformly distributed within each macrostate. 

Due to the smearing in probabilities, the CG'ing transformations constitute a semi-group \cite{fisher1998renormalization}. This irreversible character highlights a fundamental consequence of CG'ing strategies: a loss of information about the system. The definition, based on the KL divergence,  presented in Eq.~\ref{eq:smap_main} is useful for practical purposes. A more direct understanding of this information loss and how it is encoded in the mapping entropy, however, can be obtained by considering the non-ideal configurational entropies of the original and CG representation,
\begin{eqnarray}
s_r&=&-k_B\int d{\bf r}\ p_r({\bf r})\ln (V^n p_r({\bf r})) \\
s_R&=&-k_B\int d{\bf R}\ p_R({\bf R})\ln (V^N p_R({\bf R}))
\end{eqnarray}
respectively quantifying the information contained in the associated probability distributions, $p_r({\bf r})$ and $p_R({\bf R})$ \cite{shannon1948mathematical}: the higher the entropy, the more uniform the distribution, which we associate to a smaller amount of information content.
By virtue of Gibbs' inequality, from Eq.~\ref{eq:smap_main} one has $S_{map}\geq0$. Furthermore, see Appendix~\ref{app:reltomap}
\begin{equation}
\label{eq:mapent_diff}
s_R-s_r=S_{map}\geq0,
\end{equation}
so that the entropy of the CG representation is always larger than the reference, microscopic one, implying that a loss of information occurs in decreasing the level of resolution \cite{rudzinski_2011,foley2015impact}. Critically, the difference between the two information contents is precisely the mapping entropy. 

The information that is lost in the CG'ing process through $S_{map}$ only depends on the mapping operator ${\bf M}$---in our case, on the choice of the retained sites. This paves the way for the possibility of assessing the quality of a CG mapping based on the amount of information it is able to \emph{retain} about the original system, a qualitative advancement with respect to the more common a priori selection of CG representations \cite{kmiecik2016coarse}. Unfortunately, Eq.~\ref{eq:smap_main} or~\ref{eq:mapent_diff} do not allow---except in very simple cases \cite{foley2015impact}---a straightforward computational estimate of $S_{map}$ for a system arising from a choice of its CG mapping, as the observables to be averaged involve logarithms of high-dimensional probability distributions, and ultimately configuration-dependent free energies. However, having introduced the loss of information per macrostate $S_{map}({\bf R})$ defined by the relation \cite{rudzinski_2011,foley2015impact}
\begin{equation}
\label{eq:smapint}
S_{map}=\int d{\bf R}\ p_R({\bf R})S_{map}({\bf R}),
\end{equation} 
in Appendix~\ref{app:derivation} we show that this problem can be overcome by further subdividing microscopic configurations that map to a given macrostate according to their potential energy. Let us define the conditional probability $P_{\beta}(U|{\bf R})$ for the system, thermalized at inverse temperature $\beta$, to have energy $U$ provided that is in macrostate ${\bf R}$ as
\begin{eqnarray}
&&P_{\beta}(U|{\bf R})=\frac{p_R(U,{\bf R})}{p_R({\bf R})} \nonumber \\
&&=\frac{1}{p_R({\bf R})}\int d{\bf r}p_r({\bf r})\delta({\bf M}({\bf r}) - {\bf R})\delta(u({\bf r})-U),
\end{eqnarray}
so that $S_{map}({\bf R})$ can be \emph{exactly} rewritten as (see Appendix~\ref{app:derivation}):
\begin{equation}
\label{eq:smap_reweighted}
\hspace*{-0.1cm}
S_{map}({\bf R}) = k_B \ln\left[\int dU' P_{\beta}(U'|{\bf R}) \ e^{\beta(U' - \langle U\rangle_{\beta|{\bf R}})} \right],
\end{equation}
where $\langle U\rangle_{\beta|{\bf R}}$ is the average of the potential energy restricted to the CG macrostate ${\bf R}$,
\begin{equation}
\langle U\rangle_{\beta|{\bf R}}=\int dU P_{\beta}(U|{\bf R})U.
\end{equation}

This derivation enables a direct estimate of the mapping entropy $S_{map}$ from configurations sampled according to the microscopic probability distribution $p_r({\bf r})$. For a given mapping, the histogram of these configurations with respect to CG coordinates ${\bf R}$ and energy $U$ approximates the conditional probability $P_{\beta}(U|{\bf R})$ and, consequently, $S_{map}({\bf R})$, see Eq.~\ref{eq:smap_reweighted}; the total mapping entropy can thus be obtained as a weighted sum of the latter over all CG macrostates, Eq.~\ref{eq:smapint}.

The only remaining difficulty consists in obtaining accurate estimates of the exponential average in Eq.~\ref{eq:smap_reweighted}, which are prone to numerical errors. As often in these cases \cite{park2003free,chipot2007free}, it is possible to rely on a cumulant expansion of Eq.~\ref{eq:smap_reweighted}, which truncated at second order provides
\begin{equation}
\label{eq:smap_r_cum}
S_{map}({\bf R})\simeq k_B\frac{\beta^2}{2}\langle(U-\langle U\rangle_{\beta|{\bf R}})^2\rangle_{\beta|{\bf R}}.
\end{equation}
Inserting Eq. \ref{eq:smap_r_cum} in Eq.~\ref{eq:smapint} results in a \emph{total} mapping entropy given by:
\begin{eqnarray}
\label{eq:smap_cum.1_main}
S_{map} 
&\simeq& k_B \frac{\beta^2}{2} \int d{\bf R} p_R({\bf R}) \langle(U-\langle U\rangle_{\beta|{\bf R}})^2\rangle_{\beta|{\bf R}}.
\end{eqnarray}

For a CG representations to exhibit an exactly zero mapping entropy, it is required that all microstates ${\bf r}$ that map onto a given macrostate $\bf{R} = {\bf M}({\bf r})$ have the same energy in the reference system. Indeed, in this case one has $P_{\beta}(U|{\bf R})=\delta(U-\bar{u}_{\bf{R}})$ in Eq.~\ref{eq:smap_reweighted}, with $\bar{u}_{\bf{R}}$ being the potential energy common to all microstates within macrostate ${\bf R}$, and consequently $S_{map}({\bf R})=0$. Eq.~\ref{eq:smap_r_cum} highlights that deviations from this condition result in a loss of information associated to a particular CG macrostate that is proportional to the variance of the potential energy of all the atomistic configurations that map to $\bf{R}$. The overall mapping entropy is an average of these energy variances over all macrostates, each one weighted with the corresponding probability.

In the numerical implementation we thus seek to identify those mappings that cluster together atomistic configurations having the same, or at least very close energy, so as to minimise the information loss arising from CG'ing. With respect to  Eq. \ref{eq:smap_cum.1_main}, we further approximate $S_{map}$ to its discretised counterpart (see Methods),
\begin{equation}
\label{eq:tilde_smap}
\tilde{S}_{map} = k_B \frac{\beta^2}{2} \sum_{i=1}^{N_{cl}} p_R({\bf R}_i) \langle(U-\langle U\rangle_{\beta|{\bf R}_i})^2\rangle_{\beta|{\bf R}_i}
\end{equation}
where we identify $N_{cl}$ discrete CG macrostates ${\bf R}_i$, each of which contributes to $\tilde{S}_{map}$ with its own probability $p_R({\bf R}_i)$ taken as the relative population of the cluster. We then employ an algorithmic procedure to estimate and efficiently minimise, over the possible mappings, a cost function (Eq. \ref{ave_smap} of the Methods section)
\begin{equation}
\label{eq:sigma}
\asmap{} \equiv\langle \tilde{S}_{map} \rangle
\end{equation}
defined as an average of values of $\tilde{S}_{map}$ computed over different CG configuration sets, each of these being associated to a given number of conformational clusters $N_{cl}$.

\subsection{Biological structures}

It is worth stressing that the results of the previous section are completely general and independent of the specific features of the underlying system. Of course, characteristics of the input such as the force field quality, the simulation duration, the number of conformational basins explored etc. will impact the outcome of the analysis, as it is necessarily the case in any computer-aided investigation; nonetheless, the applicability of the method is not prevented or limited by these features or other system properties, e.g. the specific molecule under examination, its complexity, its size, or its underlying all-atom modelling.

To illustrate the method in its generality, we here focus our attention on three proteins we chose to constitute a small yet representative set of case studies. These molecules cover a size range spanning from $\sim 30$ to $\sim 400$ residues and a similarly broad spectrum of conformational variability and biological function, and can be taken as examples of several classes of enzymatic as well as non-enzymatic proteins.

Each protein is simulated for $200$ ns in the NVT ensemble with physiological ion concentration. Out of $200$ ns, snapshots every $20$ ps are extracted from each trajectory, for a total $10^4$ AA configurations per protein employed throughout the analysis. Details about the simulation parameters, a quantitative inspection of MD trajectories, characteristic features of each protein's results, as well as the validation of the latter with respect to the duration of the MD trajectory employed, can be found in the Supplemental Material. Hereafter we provide a description of each molecule, along with a brief summary of its behaviour as observed along MD simulations.

\noindent {\bf [TAM]} A recently released $31$-residue \emph{tamapin} mutant (PDB code 6D93). Tamapin is the toxin produced by the Indian red scorpion. It features a remarkable selectivity towards a peculiar calcium-activated potassium channel (SK2), whose potential use in the pharmaceutical context has made it a preferred object of study during the past decade \cite{Pedarzani_2002, gati_2012}. Throughout our simulation almost every residue is highly solvent-exposed. Side chains fluctuate substantially,  thus giving rise to an extreme structural variability.

\noindent {\bf [AKE]} \emph{Adenylate Kinase} (PDB code 4AKE). It is a $214$ residue-long phosphotransferase enzyme that catalyses the interconversion between adenine diphosphate (ADP) and monophosphate (AMP) and their energetically rich complex, Adenine triphosphate (ATP) \cite{4AKE}. It can be subdivided in three structural domains, CORE, LID, and NMP \cite{shapiro_2009}. The CORE domain is stable, while the other two undergo large conformational changes \cite{formoso_2015}. Its central biochemical role in the regulation of the energetic balance of the cell and relatively small size, combined with the possibility to observe conformational transitions over timescales easily accessible by plain MD \cite{ake_transitions}, make it the ideal candidate to test and validate novel computational methods \cite{wang_2020, potestio_jctc, seyler_2014}. In our MD simulation the protein displays many rearrangements in the two motile domains, which occur to be quite close at many points. Nevertheless, the protein does not undergo a full \emph{open~$\leftrightarrow$~closed} conformational transition.

\noindent {\bf [AAT]} \emph{$\alpha-1$ antitrypsin} (PDB code 1QLP). With $5934$ atoms ($372$ residues), this protein is almost two times bigger than adenylate kinase. $\alpha-1$ antytripsin is a globular biomolecule and it is well known to exhibit a conformational rearrangement over the timescales of the minutes \cite{scott_1986,nukiwa_1988,luisetti_2004}. During our simulated trajectory the molecule experiences fluctuations particularly localised in correspondence of the most solvent-exposed residues. The protein bulk appears to be very rigid, and there is no sign of a conformational rearrangement.

\subsection{Minimisation of the mapping entropy and characterisation of the solution space}

The algorithmic procedure described in the Methods section and Appendix~\ref{app:derivation} enables one to quantify the information loss experienced by a system as a consequence of a \emph{specific} decimation of its degrees of freedom. This quantification, which is achieved through the approximate calculation of the associated mapping entropy, opens the possibility of minimising such measure in the space of CG representations, so as to identify the mapping that, for a given number of CG sites $N$, is able to preserve as much information as possible about the AA reference.

In the following we allow CG sites to be located only on heavy atoms, thus reducing the maximum number of possible sites to $N_{heavy}$. We then investigate the properties of various kinds of CG mappings having different numbers of retained sites $N$. Specifically, we consider three chemically-intuitive values of $N$ for each biomolecule: (\emph{i}) $N_{\alpha}$, i.e., the number of C$_\alpha$ atoms of the structure (equal to the number of amino acids); (\emph{ii}) $N_{\alpha\beta}$, the number of C$_\alpha$ and C$_{\beta}$ atoms; and (\emph{iii}) $N_{bkb}$, which results from counting all the heavy atoms belonging to the main chain of the protein. The values of $N$ for mappings (\emph{i})-(\emph{iii}) in the case of TAM, AKE and AAT are listed in Tab.~\ref{tab:N}, together with the corresponding $N_{heavy}$. 

Even restricting $N$ to $N_{\alpha},N_{\alpha\beta}$ and $N_{bkb}$, the combinatorial dependence of the number of possible decimation mappings on the amount of retained sites and $N_{heavy}$ makes their exhaustive exploration unfeasible in practice (see Methods). To identify the CG representations that minimise the information loss we thus rely on a Monte Carlo simulated annealing approach (SA, see Methods) \cite{sa_Kirkpatrick, sa_salesman}. For each analysed protein and value of $N$, we perform $48$ independent optimisation runs, i.e., minimisations of the mapping entropy with respect to the CG site selection; we then store the CG representation characterised by the lowest value of \asmap{} in each run, thus resulting in a pool of {\it optimised} solutions. In order to assess their statistical significance and properties, we also generate a set of random mappings and calculate the associated \asmap{}'s, which constitute our reference values.

\begin{figure*}
		\includegraphics[width=\textwidth]{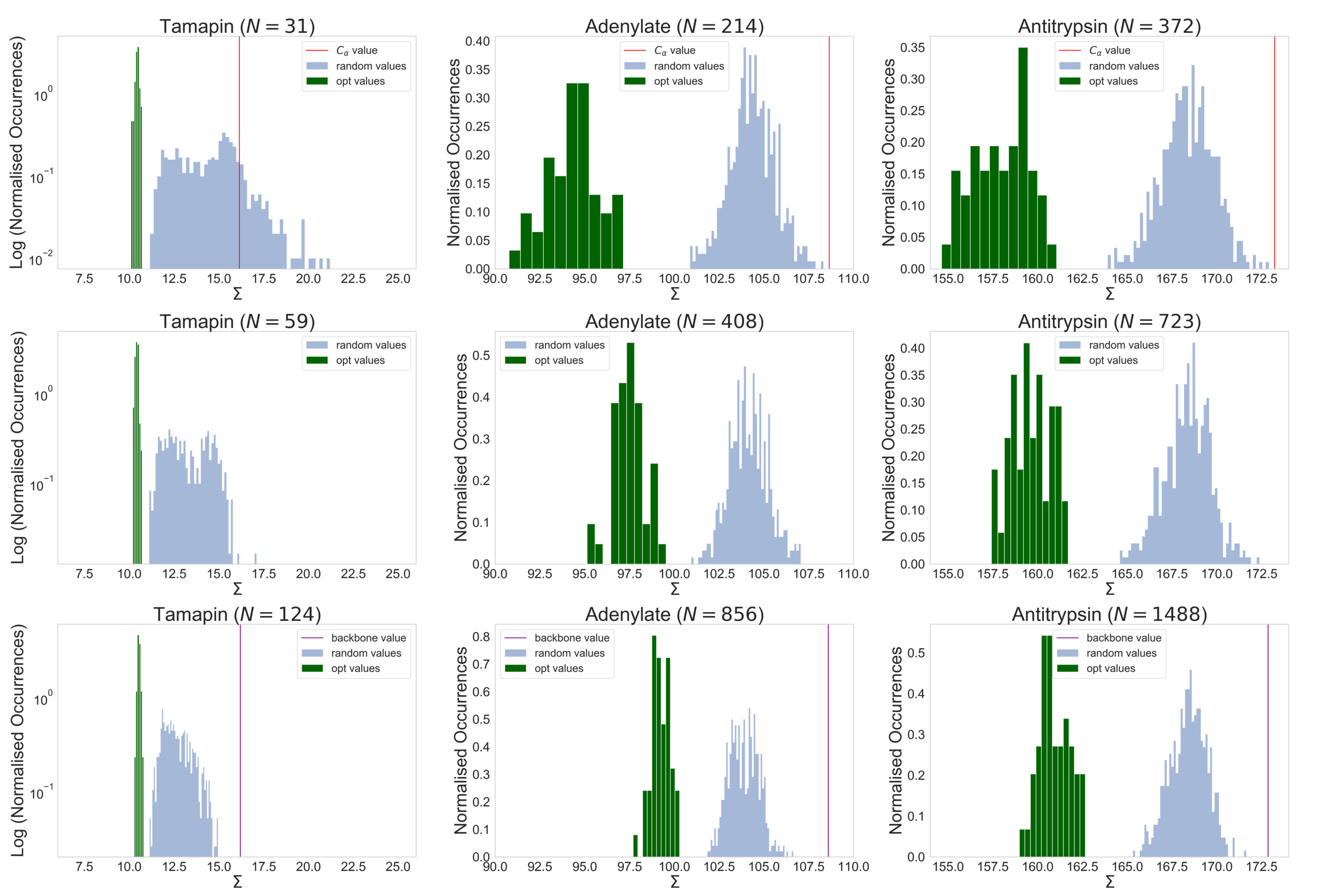}
		\caption{\label{fig:optimisations} Distributions of the values of mapping entropy $\Sigma$ [$kJ/\text{mol}/K$] in Eq.~\ref{eq:sigma} for random mappings (light blue histograms) and optimised solutions (green histograms). Each column corresponds to an analysed protein, each row to a given number $N$ of retained atoms. In the first and last rows, corresponding to numbers of CG sites equal to the number of C$_\alpha$ atoms and of backbone atoms, $N_{\alpha}$ and $N_{bkb}$ respectively, the values of the mapping entropy associated to the physically-intuitive choice of the CG sites (see text) is indicated by vertical lines (red for $N=N_{\alpha}$, purple for $N=N_{bkb}$). Note that the $S_{map}$ ranges have the same width in all plots.}
\end{figure*}

 Fig.~\ref{fig:optimisations} displays, for each value of $N$ considered, the distribution of mapping entropies obtained from a random choice of the CG representation of TAM, AKE, and AAT together with each protein's optimised counterpart. For $N=N_{bkb}$ and $N=N_{\alpha}$ in Fig.~\ref{fig:optimisations} we also report the values of \asmap{} associated to physically-intuitive choices of the CG mapping that are commonly employed in the literature: the backbone mapping ($N = N_{bkb}$), which neglects all atoms belonging to the side chains; and the C$_\alpha$ mapping ($N = N_{\alpha}$), in which we only retain the C$_\alpha$ atoms of the structures. The first is representative of united-atom CG models, while the second is a ubiquitous and rather intuitive choice to represent a protein in terms of a single bead per amino acid \cite{kmiecik2016coarse}.

\begin{table}[]
\centering
\begin{tabular}{ |P{3.7cm}||P{1.05cm}|P{1.05cm}|P{1.05cm}|P{1.05cm}|   }
 \hline
 Protein &  $N_{\alpha}$ &$N_{\alpha\beta}$&$N_{bkb}$&$N_{heavy}$\\
 \hline
 Tamapin (TAM)  & 31    &59&   124 & 230\\
 Adenylate Kinase (AKE)&   214  & 408   &856 &1656\\
 $\alpha-1$ antytripsin (AAT) & 372 & 723 & 1488 & 2956\\
 \hline
\end{tabular}
\caption{\label{table_ncg}Values of $N_{\alpha}$, $N_{\alpha\beta}$, $N_{bkb}$ and $N_{heavy}$ (see text) for each analysed protein.}\label{tab:N}
\end{table}

The optimality of a given mapping with respect to a random choice of the CG sites can be quantified in terms of the Z-score
\begin{equation}
    Z = \frac{\asmap_{opt} - \mu}{\sigma},
\end{equation}
where $\mu$ and $\sigma$ represent mean and standard deviation of the distribution of \asmap\ over randomly sampled mappings, respectively. Table~\ref{tab:zscores} summarises the values of $Z$ found for each $N$ for the proteins under examination, including $Z[backbone]$ and $Z[{\mbox{C}_{\alpha}}]$, which are computed with respect to the random distribution generated with $N = N_{bkb}$ and $N = N_{\alpha}$ respectively.

\begin{table}
\begin{flushleft}
\begin{tabular}{ |P{1.9cm}||P{2.0cm}|P{2.2cm}|P{2.0cm}| }
 \hline
 $N$ &  TAM &AKE &AAT\\
 \hline
$\overline{Z}[N_{\alpha}]$ &$-2.22 \pm 0.06$ &$-7.85 \pm 1.14$ & $-6.96 \pm 1.03$\\
$\overline{Z}[N_{\alpha\beta}]$ & $-2.38 \pm 0.08$& $-6.09 \pm 0.79$ &$-6.64 \pm 0.84$\\
$\overline{Z}[N_{bkb}]$ & $-2.65 \pm 0.09$ & $-5.55 \pm 0.62$ &$-7.24 \pm 0.85$\\
${Z}[backbone]$ & 4.37 & 5.65 & 4.31\\
${Z}[\mbox{C}_{\alpha}]$ & 0.87 & 3.36 & 3.28\\
 \hline
\end{tabular}
\caption{\label{tab:zscores} Table of $Z$ scores of each analysed protein. We report the mean and standard deviation of the distribution of $Z$ values of the optimised solutions, $\overline{Z}$, for all values of $N$ investigated. Results for the standard mappings---${Z}[backbone]$ for backbone atoms only and ${Z}[\mbox{C}_{\alpha}]$ for C$_\alpha$ atoms only---are also included.}
\end{flushleft}
\end{table}

As for the physically intuitive CG representations, Fig.~\ref{fig:optimisations} shows that the value of \asmap{} associated to the backbone mapping is very high for all structures. For TAM in particular, the amount of information retained is so low that the mapping entropy falls $4.37$ standard deviations higher than the reference distribution of random mappings, see Table~\ref{tab:zscores}. This suggests that neglecting the side chains in a CG representation of a protein is detrimental, at least as far as the structural resolution is concerned. In fact, the backbone of the protein undergoes relatively minor structural rearrangements when exploring the neighbourhood of the native conformation, thereby inducing negligible energetic fluctuations; for side chains, on the other hand, the opposite is true, with comparatively larger structural variability and a similarly broader energy range associated to it. Removing side chains from the mapping induces the clustering of atomistically different structures with different energies onto the same coarse-grained configuration, the latter being solely determined by the backbone. The corresponding mapping entropy is thus large---worse than a random choice of the retained atoms---since it is related to the variance of the energy in the macrostate.

Calculations employing the C$_\alpha$ mapping for the three structures show that this provides \asmap{} values that are very close to the ones we find with the backbone mapping, thus suggesting that C$_\alpha$ atoms retain about the same amount of information that is encoded in the backbone. This is reasonable, given the rather limited conformational variability of the atoms along the peptide chain. However, a comparison of the random case distributions for a number $N_\alpha$ and $N_{bkb}$ of retained atoms in Fig.~\ref{fig:optimisations} reveals that the former generally has a broader spread than the latter, due to the lower number of CG sites; consequently, the \asmap{} of the C$_\alpha$ atoms mapping is closer to the bulk of the distribution of the random case than that of the backbone mapping.

We now discuss the case of optimised mappings, that is, CG representations retaining the maximum amount of information about the AA reference. Each of the $48$ minimisation runs, which have been carried out for each protein in the set and value of $N$ considered, provided an optimal solution---a deep local minimum in the space of CG mappings; the corresponding \asmap{}'s spread over a compact range of values that are systematically lower than, and do not overlap with, those of the random case distributions (Fig.~\ref{fig:optimisations}).

Optimal solutions for AKE and AAT span a wide interval of values of \asmap{}; when $N = N_\alpha$ in particular, the support of this set and of the corresponding random reference have comparable sizes. A quantitative measure of this broadness is displayed in the distributions of $Z$ scores of optimal solutions presented in Table~\ref{tab:zscores}. On the other hand, TAM shows a narrower distribution of optimal values of \asmap{} for all values of $N$. As discussed in Sec.~\ref{sec:bio_sig}, this can be ascribed to the fact that most of the energy fluctuations in TAM---and consequently mapping entropy---are carried by a subset of atoms that are almost always retained in each optimal mapping, in contrast to a random choice of the CG representation. At the same time, the associated $Z$ scores are lower than the ones of the bigger proteins for all values of $N$ under examination, as TAM conformations generally feature a lower variability in energy than the other molecules.

For all the investigated proteins, the absence of an overlap between the distributions of \asmap{} associated to random and optimised mappings raises some relevant questions. First, one might wonder what kind of structure the {\it solution space} has, that is, if the identified solutions lie at the bottom of a rather flat vessel or, on the contrary, each of them is located in a narrow well, neatly separated one form the other. Second, it is reasonable to ask whether some degree of similarity exists between these quasi-degenerate solutions of the optimisation problems and, in case, what significance this has.

\begin{figure*}
		\includegraphics[width=\textwidth]{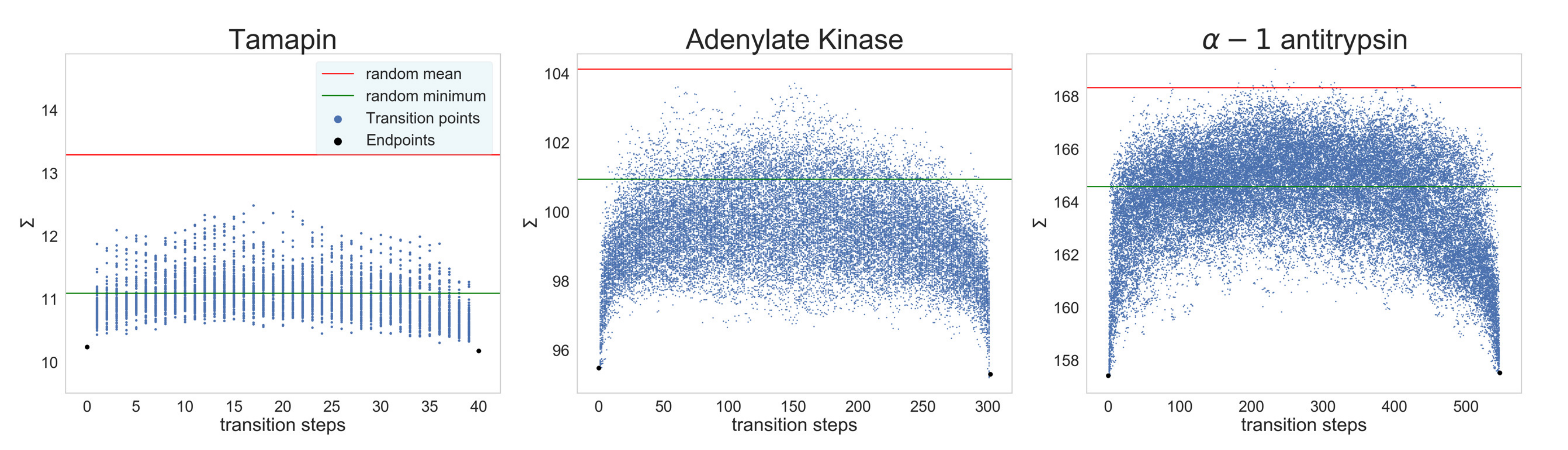}
		\caption{\label{fig:transitions} Values of the mapping entropy $\Sigma~[kJ/\text{mol}/K]$  of mappings connecting two optimal solutions. In each plot, one per protein under examination, the two lowest-$\Sigma$ mappings are taken as initial and final endpoints (black dots) for paths constructed by swapping pairs of atoms between them (blue dots). For each protein, $100$ independent paths at given $N = N_{\alpha\beta}$ are constructed and the mapping entropy of each intermediate point is computed. In each plot, horizontal lines represent the mean (red) and minimum (green) $S_{map}$ obtained from the corresponding distribution of random mappings presented in Fig.~\ref{fig:optimisations}.}
\end{figure*}

In order to answer these questions, for each structure we select four pairs of mapping operators ${\bf M}^{opt}$ that result in the lowest values of \asmap{}. We then perform $100$ independent transitions between these solutions, constructing intermediate mappings by randomly swapping two non-overlapping atoms from the two solutions at each step and calculating the associated mapping entropy. Fig.~\ref{fig:transitions} shows the results of this analysis for the pair of mappings with the lowest \asmap{}, all the other transitions being reported in Fig.~S2 of the Supplemental Material. It is interesting to notice that the endpoints (that is, the optimised mappings) correspond to the lowest values of \asmap{} along each transition path; by increasing the size of the proteins, the values of \asmap{} for intermediate mappings get closer to the average of \asmap{}$_{random}$. We cannot rule out, by this analysis, the absence of lower minima over all the possible paths, although it seems quite unlikely given the available sampling.

This analysis thus addresses the first question by showing that at least the deepest solutions of the optimisation procedure are distinct from each other. It is not possible to (quasi)continuously transform an optimal mapping into another through a series of steps keeping the value of the mapping entropy low. Each of the inspected solutions is a small town surrounded by high mountains in each direction, isolated from the others with no valley connecting them.

The second question, namely what similarity, if any, exists among these disconnected solutions, is tackled in the following section.

\subsection{Biological Significance}
\label{sec:bio_sig}
\begin{figure*}
		\includegraphics[width=\textwidth]{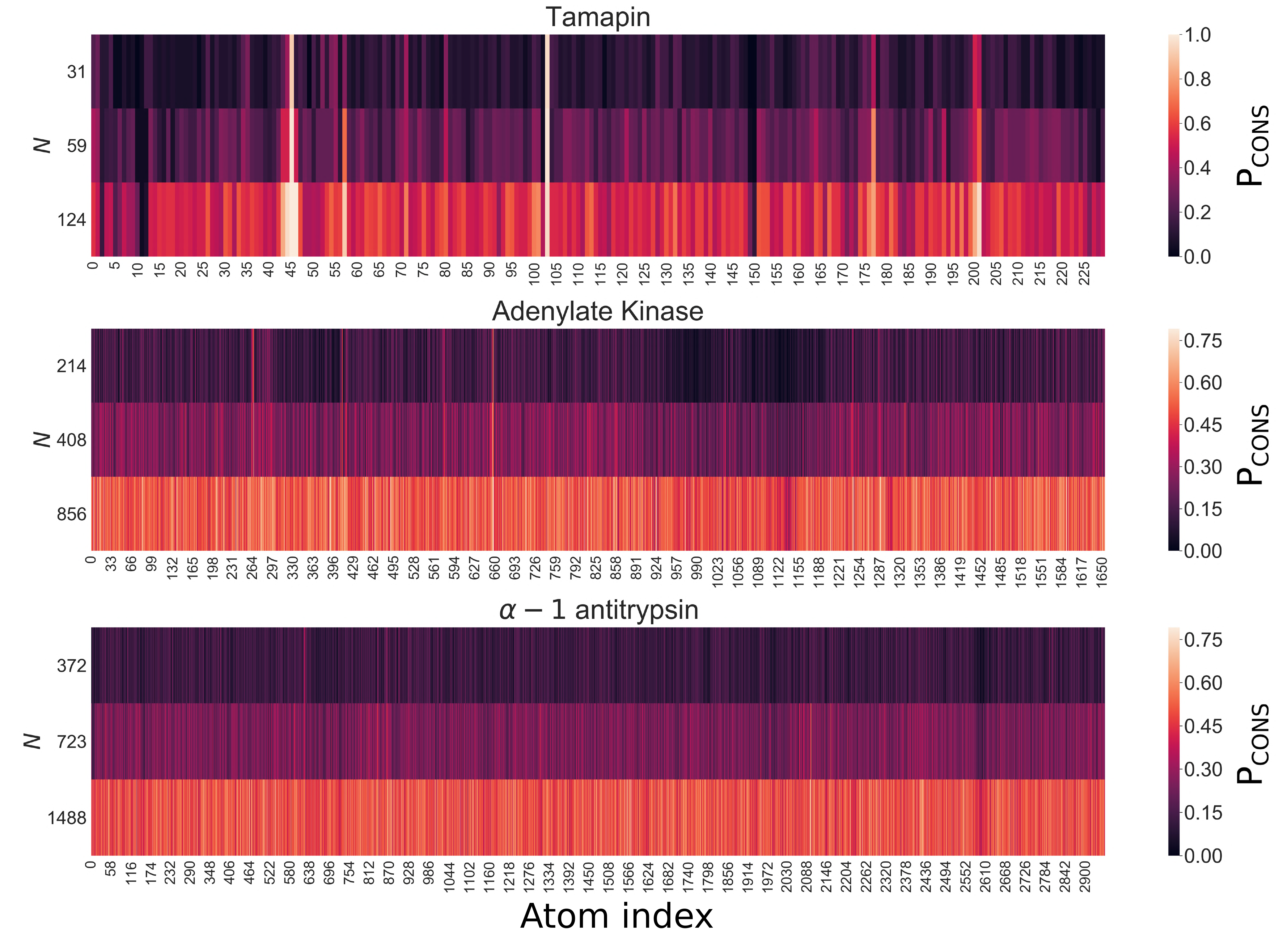}
		\caption{\label{fig:heatmap} Probability $P_{\text{cons}}$ that a given atom is retained in the optimal mapping at various numbers $N$ of CG sites and for each analysed protein, expressed as a function of the atom index. Atoms are ordered according to their number in the PDB file.}
\end{figure*}

The degree of similarity between the optimal mappings can be assessed by a simple average, returning the frequency with which a given atom is retained in the $48$ solutions of the optimisation problem.

Fig.~\ref{fig:heatmap} shows the probability $P_{\text{cons}}$ of conserving each heavy atom, separately for each analysed protein and degree of coarse-graining $N$ investigated, computed as the fraction of times it appears in the corresponding pool of optimised solutions. One can notice the presence of regions that appear to be more or less conserved. Quantitative differences can be observed between the three cases under examination: while the heat map of TAM shows narrow and pronounced peaks of conservation probability, optimal solutions for AKE feature a more uniform distribution, where the maxima and minima of $P_{\text{cons}}$ extend over secondary structure fragments rather than small sets of atoms. The distribution gets even more blurred for AAT. 

As index proximity does not imply spatial proximity in a protein structure, we mapped the aforementioned probabilities to the three-dimensional configurations. Results for TAM are shown in Fig.~\ref{fig:conf_probs_tamapin}, while the corresponding ones for AKE and AAT are provided in the Supplemental Material (Fig.~S3). From the distribution of $P_{\text{cons}}$ at different number of retained sites $N$ it is possible to infer some relevant properties of optimal mappings.

For what concerns TAM (Fig.~\ref{fig:conf_probs_tamapin}), it seems that, at the highest degree of CG ($N = N_{\alpha}$), only two sites are always conserved, namely two nitrogen atoms belonging to ARG6 and ARG13 residues ($P_{\text{cons}} (\text{NH1},\text{ARG6})$ = $0.92$, $P_{\text{cons}} (\text{NH2},\text{ARG13})$ = $0.96$). The atoms that constitute the only other arginine residue, ARG7, are well conserved but with lower probability. By increasing the resolution ($N = N_{\alpha\beta}$), i.e., employing more CG sites, we see that the atoms in the side chain of LYS27 appear to be retained more than average together with atoms of GLU24 ($P_{\text{cons}} (\text{NZ},\text{LYS27})$ = $0.65$, $P_{\text{cons}} (\text{OE2},\text{GLU24})$ = $0.75$). At $N = 124$ the distribution becomes more uniform, but still sharply peaked around terminal atoms of ARG6 and ARG13.

Interestingly, ARG6 and ARG13 have been identified to be the main actors involved in the TAM-SK2 channel interaction \cite{andreotti_2005, quintero_hernandez_2013, ramirez_cordero_2014}: Andreotti {\it et al.} \cite{andreotti_2005} suggested that these two residues strongly interact with the channel through electrostatics and hydrogen bonding. Furthermore, Ram\'irez-Cordero {\it et al.} \cite{ramirez_cordero_2014} showed that mutating one of the three arginines of TAM dramatically decreases its selectivity towards the SK2 channel.

It thus appears that the mapping entropy minimisation protocol was capable of singling out the two residues that are crucial for a complex biological process. The rationale for this can be found in the fact that such atoms strongly interact with the remainder of the protein, so that small variations of their relative coordinates have a large impact on the value of the overall system's energy. Retaining these atoms, and fixing their position in the coarse-grained conformation, thus enables the model to discriminate effectively a macrostate from another.

We note that this result was achieved solely relying on data obtained in standard MD simulations. This aspect is particularly relevant as the simulation was performed in absence of the channel, whose size is substantially larger than that of TAM. Consequently, we stress that valuable biological information, otherwise obtained {\it via} large-scale, multi-complex simulations, bioinformatic approaches, or experiments, can be retrieved by means of straightforward simulations of the molecule of interest in absence of its substrate.
 
\begin{figure*}
		\includegraphics[width=\textwidth]{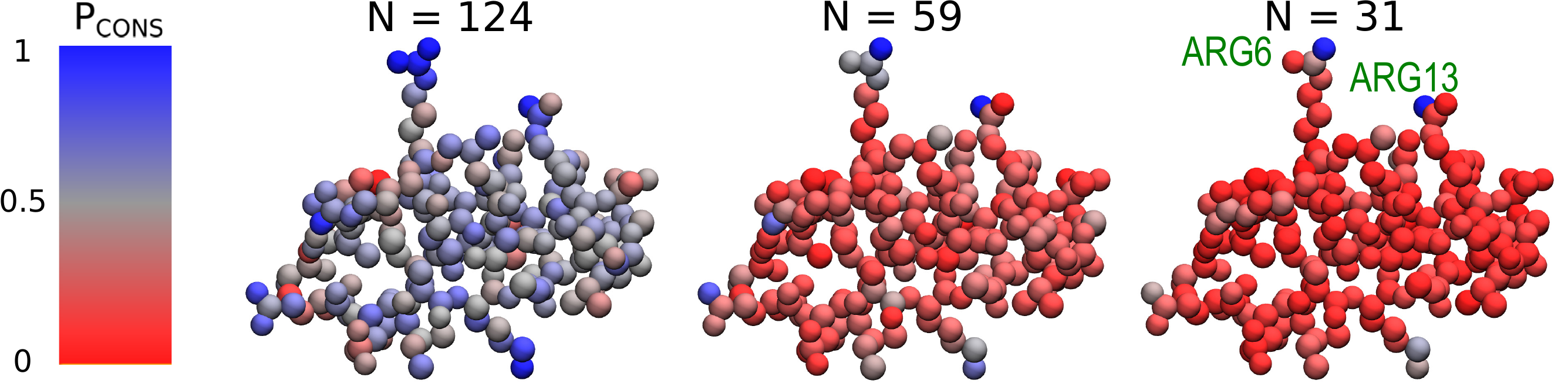}
		\caption{\label{fig:conf_probs_tamapin} Structure of tamapin (one bead per atom) coloured according to the probability $P_{\text{cons}}$ for each atom to be retained in the pool of optimal mappings. Each structure corresponds to a different number $N$ of retained CG sites. Residues presenting the highest retainment probability across $N$ (ARG6 and ARG13) are highlighted.}
\end{figure*}

For the AKE (Fig.~S3 in the Supplemental Material) we have that when $N = N_{\alpha}$ the external, solvent-exposed part of the LID domain is heavily coarse-grained, while its internal region is more conserved. The CORE region of the protein is always largely retained, without noteworthy peaks in probability. Such peaks, on the contrary, appear in correspondence of some terminal nitrogens of ARG36,  LYS57 and ARG88 ($P_{\text{cons}} (\text{NH2},\text{ARG36})$ = $0.52$, $P_{\text{cons}} (\text{NZ},\text{LYS57})$ = $0.48$, $P_{\text{cons}} (\text{NH2},\text{ARG88})$ = $0.58$). The two arginine amino acids are located in the internal region of the NMP arm, at the interface with the LID domain. ARG88 is known to be the most important residue for catalytic activity \cite{thach_2014, bellinzoni_2006}, being central in the process of phosphoryl transfer \cite{reinstein_1989}. Phenylglyoxal \cite{akbari_2012}, a drug that mutates ARG88 to a glycine, has been shown to substantially hamper the catalytic capacity of the enzyme \cite{reinstein_1989}. ARG36 is also bound to phosphate atoms \cite{bellinzoni_2006}. Finally, LYS57 is on the external part of NMP and has been identified to play a pivotal role in collaboration with ARG88 to block the release of adenine from the hydrophobic pocket of the protein \cite{matsunaga_2012}. More generally, this amino acid is crucial for stabilising the closed conformation of the kinase \cite{gur_2013, halder_2017}, which was never observed throughout the simulation. The overall probability pattern persists as $N$ increases, even though less pronounced.

As for AAT, Fig.~S3 shows that the associated optimisations heavily coarse-grain the reactive center loop of the protein. On the other hand, two of the most conserved residues in the pool of optimised mappings, MET358 and ARG101, are central to the biological role of this serpin. MET358 ($P_{\text{cons}} (\text{CE},\text{MET358})$ = $0.31$) constitutes the reactive site of the protein \cite{schapira_1986}. Being extremely inhibitor-specific, mutations or oxidation of this amino acid lead to severe diseases. In particular, heavy oxidation of MET358 is one of the main causes of emphysema \cite{taggart_2000}. The AAT \emph{Pittsburgh} variant shows MET358--ARG mutation, which leads to diminished anti-elastase activity but markedly increased antithrombin activity \cite{nejm_1983, schapira_1986, scott_1986}. In turn, ARG101 ($P_{\text{cons}} (\text{CZ},\text{ARG101})$ = $P_{\text{cons}} (\text{NH1},\text{ARG101})=P_{\text{cons}} (\text{NH2},\text{ARG101})$ = $0.35$) has a crucial role is due to its connection to mutations that lead to severe AAT deficiency \cite{luisetti_2004, nukiwa_1988}.

In summary we observe that, in all the proteins investigated, the presented approach identifies biologically relevant residues. Most notably, these residues, which are known to be biologically active in presence of other compounds, are singled out {\it from substrate-free MD simulations}. With the exception of MET358 of AAT, the most probably retained atoms belong to amino acids that are charged and highly solvent-exposed. To quantify the statistical significance of the selection operated by the algorithm, we note that the latter detects those fragments out of a pool of 8, 69 and 100 charged residues for TAM, AKE and AAT, respectively. If we account for solvent exposition, these numbers reduce to 7, 32 and 40 considering amino acids with solvent accessible surface area (SASA) higher than $1$ nm$^2$.

Another aspect worth mentioning is the fact that several atoms pinpointed as highly conserved in optimal mappings are located in the side chains of relatively large residues, such as arginine, lysine and methionine. It is thus legitimate to wonder whether a correlation might exist between an amino acid size and the probability of one or more of its atoms to be present in a low $S_{map}$ reduced representation. An inspection of the RMSF values of the three proteins' atoms {\it vs.} their conservation probability (see Fig. S4 in the Supplemental Material) shows no significant correlation for low or intermediate values of $P_{\text{cons}}$; highly conserved atoms, on the other hand, tend to be located on highly mobile residues because a relatively large conformational variability is a prerequisite for an atom to be determinant in the mapping. In conclusion, highly mobile residues are not necessarily highly conserved, while the opposite is more likely.

\section{Discussion and Conclusions}

In this work we have addressed the question of identifying the subset of atoms of a macromolecule, specifically a protein, that entails the largest amount of information about its conformational distribution while employing a reduced number of degrees of freedom with respect to the reference. The motivation behind this objective is to provide a synthetic yet informative representation of a complex system, simulated in high resolution but observed in low resolution, thus rationalising its properties and behaviour in terms of relatively few important variables--namely the positions of the retained atoms.

This goal was pursued making use of tools and concepts largely borrowed from the field of coarse-grained modelling, in particular bottom-up coarse-graining. The latter term identifies a class of theoretical and computational strategies employed to construct a simplified model of a system that, if treated in terms of a high-resolution description, would otherwise be too onerous to simulate. Coarse-graining methods make use of the configurational landscape of the reference high-resolution model to construct a simplified representation that retains its large-scale properties. The interactions among effective sites are parametrised by directly integrating out (in an exact or approximate manner) the higher-resolution degrees of freedom, and imposing the equality of the probability distributions of the coarse-grained degrees of freedom in the two representations \cite{noid_persp}.

These approaches have a long and successful history in the field of statistical mechanics and condensed matter, the most prominent, pioneering example probably being Kadanoff's spin block transformations of ferromagnetic systems \cite{kadanoff1966scaling}. This process, which lies at the heart of real-space renormalisation group (RG) theory, allows the relevant variables of the system to naturally emerge out of a (potentially infinite) pool of fundamental interactions, thus linking microscopic physics to macroscopic behaviour \cite{ma2018modern, zinn2007phase}.

The generality of the concepts of renormalisation group and coarse-graining has naturally taken them outside of their native environment \cite{schafer2012excluded,cavagna2019dynamical,antonov2017scaling}, the whole field of coarse-grained modelling of soft matter being one of the most fruitful offsprings of this cross-fertilisation \cite{noid_persp}. However, a straightforward application of RG methods in this latter context is severely restricted by fundamental differences between the objects of study. Most notably, the crucial assumptions of self-similarity and scale invariance, which justify the whole process of renormalisation at the critical point, clearly do not apply to, say, a protein, in that the latter does certainly not resemble itself upon resolution reduction. Furthermore, scaling laws cannot be applied to a system such as a biomolecule that is intrinsically finite, for which the thermodynamic limit is not defined.

Additionally, one of the key consequences of self-similarity at the critical point is that the filtering process put forward by the renormalisation group turns out to be largely independent of the specific coarse-graining prescription: the set of relevant macroscopic variables emerges as such for almost whatever choice of mapping operator is taken to bridge the system across different length scales \cite{van1993regularity}. In the case of biological matter, where the organisation of degrees of freedom is not fractal, rather hierarchical---from atoms, to residues, to secondary structure elements, and so on---the mapping operator acquires instead a central role in the ``renormalisation'' process. The choice of a particular transformation rule, projecting an atomistic conformation of a molecule to its coarse-grained counterpart, more severely implies an external---i.e. not {\it emergent}---selection of which variables are relevant in the description of the system, and which others are redundant. In this way, what should be the main outcome of a genuine coarse-graining procedure is demeaned to be one of its ingredients.

It is only recently that the central importance of the resolution distribution, i.e., the definition of the CG representation, has gradually percolated in the field of biomolecular modelling \cite{foley2015impact,potestio_jctc}. Moving away from an \emph{a priori} selection of the effective interaction sites \cite{kmiecik2016coarse}, few different strategies have been developed that rather aim at the automatic identification of CG mappings. These techniques rely on specific properties of the system under examination: examples include quasi-rigid domain decomposition \cite{Gohlke2006,ZHANG_BJ_2008,ZHANG_BJ_2009,Potestio2009,aleksiev2009,ZHANG_JCTC_2010,Sinitskiy2012,Polles2013}, or graph theory--based model construction methods that attempt at creating CG representations of chemical compounds based only on their static graph structure \cite{PONZONI20151516,depabloJCTC2019,White_2018}; other approaches aim at selecting those representations that closely match the high resolution model's energetics \cite{koehlJCTC2017,potestio_jctc}. Finally, more recent strategies rooted in the field of machine learning generate discrete CG variables by means of variational autoencoders \cite{Wang_2019}. All these methods take into account the system structure, or its conformational variability, or its energy, but none of them integrates these complementary properties in a consistent framework embracing topology, structure, dynamics, and thermodynamics.

In this context, information-theoretical measures, such as the mapping entropy \cite{Shell2008,Shell2012,rudzinski_2011,foley2015impact}, can bring novel and potentially very fruitful features \cite{lengenhaggerPRX2020}.
In fact, this quantity associates structural and thermodynamical properties, so that both the conformational variability of the system and its energetics are accurately taken into account. Making use of the advantages offered by the mapping entropy, we developed a protocol to identify, in an automated, unsupervised manner, the low-resolution representation of a molecular system that maximally preserves the amount of thermodynamical information contained in the corresponding higher-resolution model.

Furthermore, the results presented here suggest that the method may be capable of identifying not only thermodynamically consistent, but also biologically informative mappings. Indeed, a central result we reported is that those atoms consistently retained with high probability across various lowest-$S_{map}$ mappings at different CG site numbers tend to be located in amino acids that play a relevant role in the function of the three proteins under examination. Most importantly, these key residues, whose biological activity consists in binding with other molecules, have been singled out on the basis of plain MD simulations of the substrate-free molecules in explicit solvent. In general, the vast majority of available techniques for the identification of putative binding or allosteric sites in proteins rely, explicitly or implicitly, on the analysis of the interaction between the molecule of interest and its partner---be that a small ligand, another protein, or else \cite{ghersi2009easymifs,ngan2012ftsite,gervasio_ACR_2020,brady2000fast,laurie2005q,hendlich1997ligsite}. This is the case, for example, of binding site prediction servers \cite{Zhang_NAR_2012,Zhang_BioInfo_2013}, which perform a structural comparison between the target protein and those archived in a precompiled, annotated database; other bioinformatic tools make use of machine learning methods \cite{jimenez2017deepsite,Zhang_JCIM_2017,krivak2018p2rank,jendele2019prankweb}---with all pros and cons that come with the training over a possibly vast, but certainly finite dataset of known cases \cite{yang2020predicting}. To the best of our knowledge, the remaining alternative methods perform a structural analysis of the protein in search of binding pockets based on purely geometrical criteria \cite{le2009fpocket,zhu2011mspocket}. The results obtained in the present work, on the contrary, suggest that a significant fraction of biologically relevant residues, whose function is intrinsically related to the interaction with other molecules, might be identified as such from the analysis of simulations {\it in absence of the substrate}. This observation would imply that a substantial amount of information about functional residues, even those that exploit their activity through the interaction with a partner molecule, is entailed in the protein's own structure and energetics. In the past few decades, the successful application of extremely simplified representations of proteins such as elastic network models has shown that the key features of a protein's large-scale dynamics are encoded in its native structure \cite{Amadei1993,Tirion1996,Bahar1997,Hinsen1998,Atilgan2001,Delarue2002,Micheletti2004,Pontiggia2007,Potestio2009,Hensen2012,Nussinov2014,Wei2016}; in analogy with this, we hypothesise that the mapping entropy minimisation protocol is capable of bringing to light those {\it relational} properties of proteins---namely the interaction with a substrate---from the thermodynamics of the single molecule, in absence of its partner.

The mapping entropy minimisation protocol establishes a quantitative bridge between a molecule's representation---and hence its information content---on the one side, and the structure-dynamics-function relationship on the other. This method might represent a novel and useful tool in various fields of applications, e.g. for the identification of important regions of proteins, such as druggable sites and allosteric pockets, relying on simple, substrate-free MD simulations, and efficient analysis tools. In this study, a first exploration of the method's capabilities, limitations, and potential developments has been carried out, and several perspectives lie ahead that deserve further exploration. Among the most pressing and interesting ones we mention the investigation of how the optimised mappings depend on the conformational space sampling; the relation of the mapping entropy minimisation with more established schemes such as the maximum entropy method; and the viability of a machine learning-based implementation of the protocol, e.g. making use of deep learning tools that have proven to be strictly related to coarse-graining, dimensionality reduction, and feature extraction. All these avenues are the object of ongoing study.

In conclusion, it is our opinion that the proposed automated selection of coarse-grained sites entails a great potential for further development, being at the nexus between molecular mechanics, statistical mechanics, information theory, and biology.

\section{Methods}

In this section we describe the technical preliminaries and the details of the algorithm we employ to obtain the CG representation ${\bf M}$, see Eq.~\ref{eq:decimation}, that minimises the loss of information inherently generated by a CG'ing procedure---that is, the mapping entropy.

Eq.~\ref{eq:smap_cum.1_main} provides us with a way of measuring the mapping entropy of a biomolecular system associated to any particular choice of decimation of its atomistic degrees of freedom. One can visualise a decimation mapping (Eq. \ref{eq:decimation}) as an array of bits, where 0 and 1 correspond to \emph{not retained} and \emph{retained} atoms, respectively. Order matters: swapping two bits produces a different mapping operator. Applying this procedure, one finds that the total number of possible CG representations of a biomolecule, irrespectively of how many atoms $N$ are selected out of $n$, is
\begin{equation}\label{eq:binomial}
\sum_{N = 0}^{n} \frac{n!}{N! \ (n-N) !} = 2^n,
\end{equation}
which is astronomical even for the smallest proteins. In this work we restrict the set of possibly retained sites to the $N_{heavy}$ heavy atoms of the compound---excluding hydrogens---thus significantly reducing the cardinality of the space of mappings. Nonetheless, finding the global minimum of Eq. \ref{eq:smap_cum.1_main} for a reasonably large molecule would be computationally intractable whenever $N$ is different from $1,2$ and $N_{heavy}-1, N_{heavy}-2$. As an example, there are $2.4 \times 10^{38}$ CG representations of tamapin with 31 sites ($N = N_{\alpha}$) and $9.6 \times 10^{887}$ for antitrypsin with 1488 sites ($N = N_{bkb}$).

Hence, it is necessary to perform the minimisation of the mapping entropy through a Monte Carlo-based optimisation procedure, and we specifically rely on the simulated annealing (SA) protocol \cite{sa_Kirkpatrick, sa_salesman}. As it is typically the case with this method, the computational bottleneck consists in the calculation of the observable (the mapping entropy) at each SA step.

We develop an approximate method that is able to obtain the mapping entropy of a biomolecule by analysing a MD trajectory that can contain up to tens of thousands of frames. At each SA step, that is, for each putative mapping, the algorithm calculates a similarity matrix among all the generated configurations. The entries of this matrix are given by the root mean square deviation (RMSD) between structure pairs, the latter being defined only in terms of the retained sites associated to the CG mapping, and aligned accordingly; we then identify CG macrostates by clustering frames based on the distance matrix, making use of bottom-up hierarchical clustering (UPGMA \cite{upgma_original}). Finally, we determine the observable of interest from the variances of the atomistic intramolecular potential energy of the protein corresponding to the frames that map onto the same CG conformational cluster, see Eq.~\ref{eq:tilde_smap}.

\begin{figure*}
		\includegraphics[width=\textwidth]{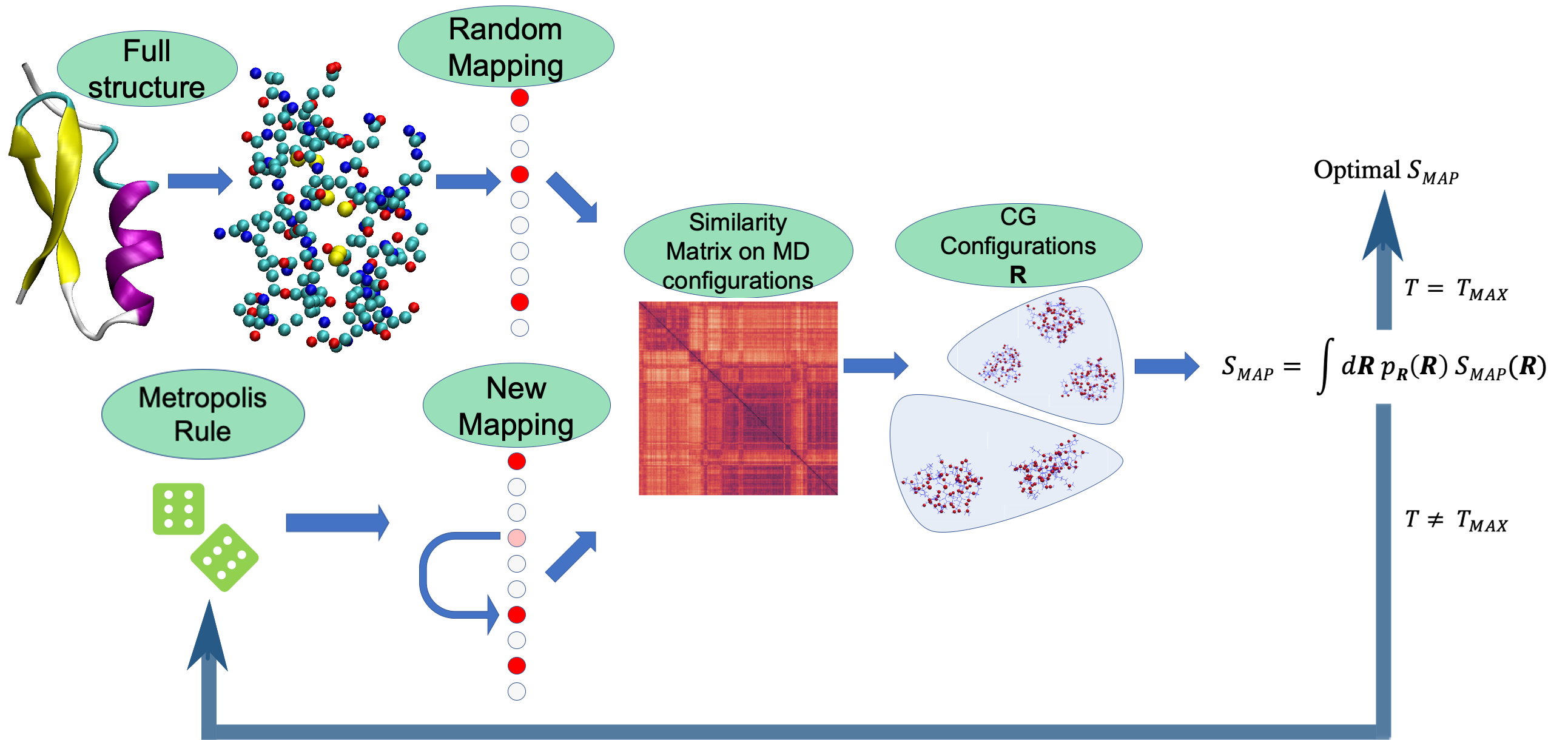}
		\caption{\label{fig:algo}Schematic representation of the algorithmic procedure described in the text that we employ to minimise the mapping entropy, the latter being calculated by means of Eq. \ref{ave_smap}. The full similarity matrix is computed once every $T_{K}$ steps, while in the intermediate steps we resort to the approximation given by Eq. \ref{eq:msd}. $T_{K}$ depends both on the protein and on $N$. $T_{MAX}$ is the number of simulated annealing steps, $T_{MAX}=2 \times 10^4$.}
\end{figure*}

The protocol is initiated with the generation of a mapping such that the overall number of retained sites is equal to $N$. Then, at each SA step, the following operations are performed:

\begin{enumerate}
	\item swap a retained site ($\sigma_i = 1$) and a removed site ($\sigma_j = 0$) in the mapping;
	\item compute a similarity matrix among CG configurations using the RMSD;
	\item apply a clustering algorithm on the RMSD matrix in order to identify the CG macrostates $\bf{R}$;
	\item compute $\tilde{S}_{map}$ using Eq. \ref{eq:tilde_smap}.
\end{enumerate}

Once the new value of $\tilde{S}_{map}$ is obtained, the move is accepted/rejected using a Metropolis-like rule. The overall workflow of the algorithm is schematically illustrated in Fig. \ref{fig:algo}.

For the sake of the accuracy of the optimisation, the more exhaustive the sampling the better, hence the number of sampled atomistic configurations should be at least of the order of the tens of thousands. However, in that case step 2 would require to align a huge number of structure pairs for each proposed CG mapping, which in turn would dramatically slow down the entire process. This problem is circumvented performing a reasonable approximation in the calculation of the CG RMSD matrix.

\subsection{RMSD matrix calculation}\label{RMSD_sec}

The RMSD between two \emph{superimposed} structures $\bm{x}$ and $\bm{y}$ is given by
\begin{equation}
    RMSD (\bm{x} , \bm{y}) = \sqrt{\frac{1}{n} \sum_{i=1}^{n} (\bm{x}_{i} - \bm{y}_{i})^{2}},
\end{equation}

where $n$ is the number of sites in the system, being they atomistic or CG, and $\bm{x}_{i}$, $\bm{y}_{i}$ represent the cartesian coordinates of the $i$-th element in the two sets. According to Kabsch \cite{Kabsch76,Kabsch78} it is possible to find the superimposition that minimises this quantity, namely the rotation matrix $U$ that has to be applied to $\bm{x}$ for a given $\bm{y}$ in order to reach the minimum of the RMSD.

The aforementioned procedure is not computationally heavy \emph{per se}; in our case, however, we would have to repeat this alignment for all configuration pairs in the MD trajectory every time a new CG mapping is proposed along the Monte Carlo process, thus making the overall workflow inctractable in terms of computational investment.

The simplest solution to this problem is to discard the differences in the Kabsch alignment between two CG structures differing by a pair of swapped atoms. This assumption is particularly appealing from the point of view of speed and memory, since the expensive and relatively slow alignment procedure produces a result (a rotation matrix) that can be stored with negligible use of resources. In order to take advantage of this simplification without losing accuracy, for each structure and degree of CG we select an interval of Simulated Annealing steps $T_{K}$ in which we consider rotation matrices constant. After these steps, the full Kabsch alignment is applied again.

This approximation results in a substantial reduction of the number of operations that we have to execute at each Monte Carlo step. At first, given the initial random mapping operator ${\bf M}$, we build the sets of coordinates that have been conserved by the mapping operator $\Gamma({\bf M}) = {\bf M}({\bf r})$. Then we compute the overall RMSD matrix between every pair of aligned structures $\Gamma_{\alpha}$ and $\Gamma_{\beta}$, $RMSD(\Gamma_{\alpha}({\bf M}), \Gamma_{\beta}({\bf M}))$, where $\alpha$ and $\beta$ run over the MD configurations. For all moves ${\bf M}\rightarrow {\bf M}'$ within a block of $T_K$ Monte Carlo steps, ${\bf M}$ and ${\bf M}'$ only differing in a pair of swapped atoms, this quantity is then updated with the simple rule
\begin{eqnarray}\label{eq:msd}
&& MSD(\Gamma_{\alpha}({\bf M}'),\Gamma_{\beta}({\bf M}')) = MSD(\Gamma_{\alpha}({\bf M}), \Gamma_{\beta}({\bf M})) - \nonumber \\ 
&&\frac{1}{N} MSD(\Gamma_{\alpha}(\bm{s}), \Gamma_{\beta}(\bm{s})) 
+ \frac{1}{N} MSD(\Gamma_{\alpha}(\bm{a}), \Gamma_{\beta}(\bm{a})),
\end{eqnarray}
where $\bm{s}$ and $\bm{a}$ are the removed (substituted) and added atom, respectively, and MSD is the Mean Squared Deviation.

This approach clearly represents an approximation to the correct procedure; it has to be emphasised, however, that the impact of such approximation is increasingly perturbative as the size of the system grows. Furthermore, the computational gain that the described procedure enables is sufficient to counterbalance the fact that the exact protocol would be so inefficient to make the optimisation impossible. For example, choosing $T_{K} = 1000$ for AAT with $N = N_{bkb}$ our approximation gives a speed-up factor of the order of $10^3$.

\subsection{Hierarchical Clustering of Coarse Grained configurations}

Several clustering algorithms exist that have been applied to group molecular structures based on RMSD similarity matrices \cite{rmsd_conformers_clust,rmsd_soms_clust}. Many such algorithms have been developed and incorporated in the most common libraries for data science. Among the various available methods we choose to resort on the agglomerative bottom-up hierarchical clustering with average linkage (UPGMA algorithm \cite{upgma_original}). We here briefly recapitulate the basics underpinnings of this procedure.

\begin{enumerate}
    \item At the first step, the minimum of the similarity matrix is found and the two corresponding entries $x,y$ (\emph{leaves}) are merged together in a new cluster $k$;
    \item $k$ is placed in the middle of its two constituents. The distance matrix is updated to take into account the presence of the new cluster in place of the two \emph{close} structures: $d(k,z) = (d(x,z) + d(y,z))/2$;
   \item Steps 1. and 2. are iterated until one \emph{root} is found. The distance among clusters $k$ and $w$ is generalized as follows:
   \begin{eqnarray}
   d(k,w) = \sum_{i \in k} \sum_{j \in w} \frac{d(k[i],w[j])}{|k|\times |w|}
    \end{eqnarray}
    where $|k|$ and $|w|$ are the populations of the clusters and $k[i]$ and $w[j]$ their elements;
    \item The actual division in clusters can be performed by cutting the tree (\emph{dendrogram}) using a threshold value on the inter-clusters distance or taking the first value of distance that gives rise to a certain number of clusters $N_{cl}$. In both cases it is necessary to introduce a hyperparameter. In our case the latter is a more viable choice to reduce the impact of roundoff errors. Indeed, the first criterion would push the optimisation to create as many clusters as possible, in order to minimise the energy variance inside them (a cluster with one sample has zero variance in energy).
\end{enumerate}

This algorithm, whose implementation \cite{scipy_millner, scipy_joseph} is available in Python Scipy \cite{scipy_std}, is simple, relatively fast ($O(n^2 \, log \, n)$), and completely deterministic: given the distance matrix, the output dendrogram is unique. 

Although this algorithm scales well with the size of the dataset, it may not be robust with respect to small variations along the optimisation trajectory. In fact, even the slightest modifications of the dendrogram may lead to abrupt changes in $\tilde{S}_{map}$. This is perfectly understandable from an algorithmic point of view, but it is deleterious for the stability of the optimisation procedure. Furthermore, the aforementioned choice of $N_{cl}$ is somehow arbitrary. Hence, we perform the following analysis in order to enhance the robustness of $\tilde{S}_{map}$ at each MC move and to provide a quantitative criterion to set the hyperparameter:

\begin{enumerate}
    \item Compute the RMSD similarity matrix between all the heavy atoms of the biological system under consideration;
    \item Apply UPGMA algorithm to this object, retrieving the all-atom dendrogram;
    \item Impose lower and upper bounds (see Table \ref{tab:ncl}) on the inter-clusters distance depending on the conformational variability of the structure;
    \item Visualise the cut dendrogram to identify the number of different clusters available at each of the two values of the threshold ($N_{cl}^+$ and $N_{cl}^-$) (Table \ref{tab:ncl});
    \item Build a list $\text{CL}$ of five integers selecting three (intermediate) values between $N_{cl}^-$ and $N_{cl}^+$;
    \item Define the observable as the average over the values of $\tilde{S}_{map}$ (see Eq.~\ref{eq:tilde_smap}) computed choosing different $N_{cl}$:
    \begin{eqnarray}
    \label{ave_smap}
    &&\asmap = \frac{1}{|\text{CL}|} \sum_{N_{cl} \in \text{CL}} \tilde{S}_{map}(N_{cl})
    \end{eqnarray}
where $|\text{CL}|$ is the cardinality of the list we chose.
\end{enumerate}

 \begin{table}
  \begin{tabular}{ |P{2.7cm}||P{1.1cm}|P{1.1cm}|P{1.1cm}|P{1.1cm}|}
 \hline
 Protein &  Upper bound (nm) & Lower bound (nm) &  $N_{cl}^+$  & $N_{cl}^-$\\
 \hline
 Tamapin   &  0.20    &  0.18 &  91 &  34\\\
 Adenylate Kinase&   0.25  &  0.20  &   147  &   29 \\
 $\alpha-1$ antytripsin & 0.20 & 0.15 & 96 &   7\\
 \hline
\end{tabular}
\caption{\label{tab:ncl}Bounds on inter-clusters distance and correspondent number of clusters.}
\end{table}

The overall procedure amounts at identifying many different sets of CG macrostates ${\bf R}$ on which $\tilde{S}_{map}$ can be computed, assuming that the average of this quantities can be used effectively as driving observable inside the optimisation. Noteworthy, this trivial assumption allows to increase the robustness of the SA optimisation and to keep in memory all the values of $\tilde{S}_{map}$ calculated at different distances from the root of the dendrogram. 

\subsection{Simulated Annealing}

We use Monte Carlo simulated annealing to stochastically explore the space of the possible decimation mappings associated to each degree of CG'ing. We here briefly describe the main features of our implementation of the SA algorithm, referring the reader to a few excellent reviews for a comprehensive description of the techniques that can be employed in the choice of temperature decay and parameter estimation \cite{PARK_syst_proc_SA,Connolly_SA_QAP}.

We run the optimisation for $2 \times 10^3$ MC epochs, each of which is composed by 10 steps. This amounts at keeping the temperature constant for 10 steps and then decreasing it according to an exponential law. For the $i$-th epoch we have that $T(i) = T_0\ e^{-i/\nu}$.

The hyperparameters $T_0$ and $\nu$ are crucial for a well-behaved MC optimisation. We choose $\nu = 300$ so that the temperature at $i = 2000$ is approximately $T_0 / 1000$. In order to feed our algorithm with reasonable values of $T_0$, for each of 100 random mappings we perform 10 MC stochastic moves, measuring $\Delta$\asmap{}, namely the difference between the observables computed at two consecutive steps. Then we estimate $T_0$ so that a move that leads to an increment of the observable equal to the average of $\Delta$\asmap{} would possess an acceptance probability of 0.75 at the first step.

\subsection{Data available}
For each analysed protein, the raw data about all the CG representations investigated in this work including random, optimised and transition mappings are freely available on the Zenodo repository \url{https://zenodo.org/record/3776293} together with the associated mapping entropies. We further provide all the scripts we employed to analyse such data and construct all the figures presented in this work.

\begin{acknowledgments}
The authors thank Attilio Vargiu for critical reading of the manuscript and useful comments. This project has received funding from the European Research Council (ERC) under the European Union's Horizon 2020 research and innovation programme (grant agreement No 758588). MSS acknowledges funding from the U.S. National Science Foundation through award CHEM-1800344.
\end{acknowledgments}

\section{Supporting information}
The Supporting Information file contains detailed information on the following topics:

\begin{itemize}
\item a quantitative analysis of the all-atom MD simulations of the three proteins investigated in this work
\item additional figures about the CG representations that minimise the mapping entropy
\item an analysis of the relation between the size and mobility of residues and the conservation probability of their atoms
\item an assessment of the results' stability with respect to the duration of the MD trajectory.
\end{itemize}

\begin{figure}[h]
\centering
\includegraphics[width=\columnwidth]{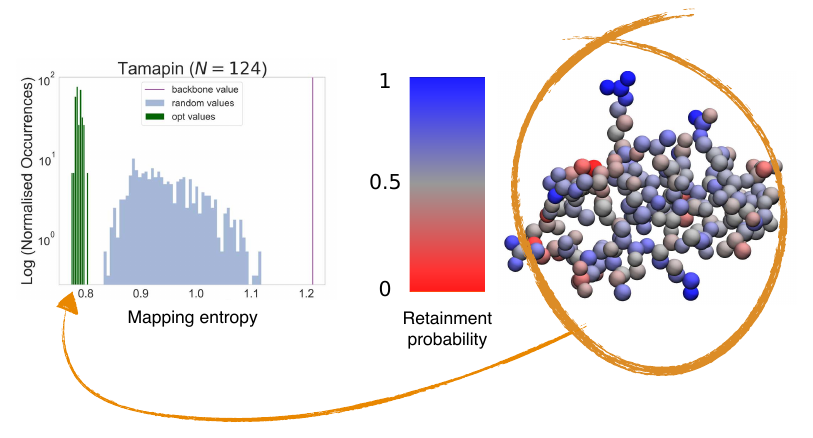}
\caption{Table of Content figure.}
\end{figure}

\appendix

\section{Relative and mapping entropy}
\label{app:reltomap}

Bottom-up coarse-graining approaches aim at constructing effective, low-resolution representations of a system that reproduce as accurately as possible the equilibrium statistical mechanical properties of the underlying, high-resolution reference. In particular, this problem is phrased in terms of the parametrisation of a CG potential that approximates the reference system's multi-body potential of mean force (PMF) $U^0$,
\begin{equation}
\label{eq:pmffull}
    U^0 = -k_B T \ln (V^Np_R({\bf R}))+const,
\end{equation}
where $p_R({\bf R})$ is the probability for the atomistic model to sample a specific CG configuration ${\bf R}$. In the canonical ensemble, one has
\begin{eqnarray}
\label{eq:pmacro}
 p_R({\bf R})&=&\int d{\bf r}\ p_{r}({\bf r})\delta({\bf M}({\bf r}) - {\bf R}) \nonumber \\
 &=&\frac{1}{Z}\int d{\bf r}\ e^{-\beta u({\bf r})}\delta({\bf M}({\bf r}) - {\bf R}),
\end{eqnarray}
where $\beta=1/k_BT$, $u({\bf r})$ is the microscopic potential energy of the system, $p_r({\bf r})\propto\exp(-\beta u({\bf r}))$ is the Boltzmann distribution and $Z$ the associated configurational partition function.

From Eqs.~\ref{eq:pmffull} and~\ref{eq:pmacro} it follows that a computer simulation of the low-resolution system performed with the potential $U^0$ (more precisely, a free energy) would allow the CG sites to sample their configurational space with the same probability as they would do in the reference system. Unfortunately, the intrinsically multi-body nature of $U^0$ is such that its exact determination is largely unfeasible in practice \cite{dijkstra1999phase}. Considerable effort has thus been devoted to devise increasingly accurate methods to approximate the PMF with a CG potential $U$ \cite{noid2008multiscale,Shell2008,rudzinski2015generalized,menichetti2017thermodynamics}; however, the latter is in general defined in terms of a necessarily incomplete set of basis functions \cite{Takada2012,noid_persp,Saunders2013,Potestio2014}. It is thus natural to look for quantitative measures of a CG model's quality with respect to $U^0$.

In this respect, one of the most notable examples of such metrics is the relative entropy \cite{Shell2008,Shell2012,rudzinski_2011,foley2015impact},
\begin{eqnarray}
\label{eq_srel}
    S_{rel} = k_B\times D_{KL}(p_{r}({\bf r})||P_{r}({\bf r}|U)) \nonumber \\ 
    = k_B\int d{\bf r}\ p_r({\bf r}) \ln \left[ \frac{p_r({\bf r})}{P_r({\bf r}|U)} \right],
\end{eqnarray}
where $D_{KL}(\cdot || \cdot)$ denotes the Kullback-Leibler divergence between two probability distributions \cite{kullback1951information}, with $S_{rel}\geq 0$ by virtue of Gibbs' inequality. In Eq.~\ref{eq_srel}, $p_{r}({\bf r})$ is the atomistic probability distribution of the system, see Eq.~\ref{eq:pmacro}, while $P_{r}({\bf r}|U)$ is defined as a product of probabilities over CG and AA configurational spaces \cite{rudzinski_2011, foley2015impact},
\begin{equation}
\label{eq:app_bigP}
P_{r}({\bf r}|U) = \frac{p_{r}({\bf r})}{p_R({\bf M}({\bf r}))} P_R({\bf M}({\bf r})|U).
\end{equation}

The term $P_R({\bf R}|U)\propto \exp(-\beta U({\bf R}))$ in Eq.~\ref{eq:app_bigP} runs over CG configurations, and describes the probability that a CG model with approximate potential $U({\bf R})$ samples the CG configuration ${\bf R}$. Then, to obtain $P_{r}({\bf r}|U)$ it is sufficient to multiply $P_R({\bf R}|U)$ by the atomistic probability $p_r({\bf r})$ of sampling ${\bf r}$, normalised by the Boltzmann weight $p_R({\bf R})$ of the CG configuration ${\bf R}$ (see Eq.~\ref{eq:pmacro}).

KL divergences quantify the information loss between probability distributions; specifically, $D_{KL}(s({\bf r}) || t({\bf r}))$ represents the information that is lost by representing a system originally described by a probability distribution $s({\bf r})$ through a distribution $t({\bf r})$ \cite{kullback1951information}. Given a CG mapping ${\bf M}$, the relative entropy $S_{rel}$ in Eq.~\ref{eq_srel} implicitly measures the loss that arises as a consequence of approximating the CG potential of mean force $U^0$ of a system by an effective potential $U$. By replacing Eq~\ref{eq:app_bigP} in Eq.~\ref{eq_srel} and introducing $1=\int d{\bf R}\ \delta({\bf M}({\bf r}) - {\bf R})$, one indeed obtains
\begin{equation}\label{eq:srelR}
S_{rel}=k_B\int d{\bf R}\ p_R({\bf R})\ln\left[\frac{p_R({\bf R})}{P_R({\bf R}|U)}\right],
\end{equation}
that is, a KL divergence $D_{KL}(p_R({\bf R})||P_R({\bf R}|U))$ between the \emph{exact} and \emph{approximate} probability distributions in the CG configuration space, with no explicit connection to the underlying microscopic reference. However, it is possible to expand $S_{rel}$ as a difference between two information losses (the one due to $U$ and the one due to $U^0$) calculated with respect to the atomistic system,
\begin{eqnarray}
\label{eq:srel_twoterms}
    S_{rel} &=&  k_B\times D_{KL}(p_r({\bf r})||V^{N-n}P_R({\bf M}({\bf r})|U)) \nonumber \\
   &-&   k_B\times D_{KL}(p_r({\bf r})||V^{N-n}p_R({\bf M}({\bf r}))) \nonumber \\
    &=& k_B \int d{\bf r}\ p_r({\bf r}) \ln \left[ \frac{V^n}{V^N}  \frac{p_{r}({\bf r})}{P_R({\bf M}({\bf r})|U)} \right]  \nonumber \\ 
    &-&  k_B\int d{\bf r}\ p_r({\bf r})\ln \left[ \frac{V^n}{V^N} \frac{p_r({\bf r})}{p_R({\bf M}({\bf r}))} \right],
\end{eqnarray}
where $n$ and $N$ denote the number of atomistic and CG sites, respectively.

Both KL divergences in Eq.~\ref{eq:srel_twoterms} are positive defined due to Gibbs' inequality; the second one is called mapping entropy \footnote{In this work we employ a different sign convention for the mapping entropy $S_{map}$ with respect to Refs.~\cite{rudzinski_2011,foley2015impact}, and consistent with the one in Ref.~\cite{Shell2008}. On one hand, this enables the mapping entropy to be directly related to a loss of information in the KL sense---a \emph{positive} KL divergence implies a \emph{loss} of information. On the other hand, it allows the relative entropy in Refs.~\cite{rudzinski_2011,foley2015impact} to be considered a difference of information losses---those of $U$ and $U^0$, see  Eq.~\ref{eq:srel_twoterms}---calculated with respect to the atomistic system, so that the vanishing of $S_{rel}$ for $U=U^0$ in Refs.~\cite{rudzinski_2011,foley2015impact} effectively amounts at recalibrating the zero of the relative entropy as originally defined in Ref.~\cite{Shell2008}.} $S_{map}$ \cite{Shell2008,rudzinski_2011,foley2015impact},
\begin{equation}
\label{eq:app_smaponeterm}
S_{map}=k_B\int d{\bf r}\ p_r({\bf r})\ln \left[ \frac{V^n}{V^N} \frac{p_r({\bf r})}{p_R({\bf M}({\bf r}))} \right]\geq0,
\end{equation}
which noteworthy does not depend on the CG force field $U$ but only on the mapping operator ${\bf M}$.

In multi-scale modelling applications, one seeks to minimise the relative entropy with respect to coefficients in terms of which the coarse-grained potential $U({\bf R})$ is parametrised \cite{Shell2008,rudzinski_2011,Shell2012,foley2015impact}. The aim is to generate CG configurations that sample the {\it atomistic} conformational space with the same microscopic probability $p_r({\bf r})$, see Eq. \ref{eq_srel}. However, since the model can only generate configurations in the CG space, minimising Eq. \ref{eq_srel} is tantamount to minimise Eq. \ref{eq:srelR}; furthermore, in the minimisation with respect to $U$ the contribution of the mapping entropy vanishes, because the latter does not depend on the coarse-grained potential. In this context, then, $S_{map}$ only represents a constant shift of the KL distance between the all-atom and the coarse-grained models, and a minimisation of the first term of Eq. \ref{eq:srel_twoterms} is equivalent to that of Eq. \ref{eq_srel}.

When taken {\it per se}, on the other hand, the mapping entropy provides substantial information about the modelling of the system. In fact, this quantity represents the loss of information that would be inherently generated by reducing the resolution of a system even in the case of an \emph{exact} CG'ing procedure, in which $U=U^0$ and $S_{rel}=0$ \cite{rudzinski_2011}. In the calculation of $S_{map}$, the reference AA density is compared to a distribution in which probabilities are smeared out and redistributed equally to all the microscopic configurations $\bf r$ inside each CG macrostate.

Starting from Eq.~\ref{eq:app_smaponeterm}, Rudzinski \emph{et al.} further divide $S_{map}$ into a sum of two terms \cite{rudzinski_2011},
\begin{eqnarray}
\label{eq:smap_twoterms}
S_{map} &=& -k_B \int d{\bf r}\ p_r({\bf r}) \ln  \left[ \frac{V^N}{V^n} \Omega_1({\bf M}({\bf r})) \right]  \nonumber \\ 
&+& k_B \int d{\bf r}\ p_r({\bf r}) \ln \left[ \frac{p_r({\bf r})}{\bar{p}_r({\bf r})} \right],
\end{eqnarray}
where the first one is purely geometrical while the second one accounts for the smearing in probability generated by the CG'ing procedure.
In Eq.~\ref{eq:smap_twoterms}, $\Omega_1({\bf M}({\bf r})) =   \int d{\bf r} \delta({\bf M}({\bf r}) - {\bf R})$ is the degeneracy of the CG macrostate ${\bf R}$---i.e., how many microstates map onto a given CG configuration---and
\begin{equation}
\label{eq:pbar_app}
\bar{p}_r({\bf r}) = {p_R({\bf M}({\bf r}))}/{\Omega_1({\bf M}({\bf r}))}
\end{equation}
is the average probability of all microstates that map to the macrostate ${\bf R}={\bf M}(\bf r)$.

The geometric term in Eq.~\ref{eq:smap_twoterms} does not vanish in general \cite{rudzinski_2011}. However, if the mapping takes the form of a decimation, see Eq.~\ref{eq:decimation}, one has
\begin{equation}\label{eq:omega1dec}
\Omega_1({\bf M}({\bf r})) = V^{n-N},
\end{equation}
and the first logarithm in Eq. \ref{eq:smap_twoterms} is identically zero, so that  

\begin{equation}
\label{eq:smap_app}
S_{map}=k_B \int d{\bf r}\ p_r({\bf r}) \ln \left[ \frac{p_r({\bf r})}{\bar{p}_r({\bf r})} \right].
\end{equation}

In the case of decimation mappings, moreover, a direct relation holds between the mapping entropy $S_{map}$ as expressed in Eq.~\ref{eq:smap_app} and the non-ideal configurational entropies of the original and CG systems \cite{rudzinski_2011,foley2015impact},
\begin{eqnarray}
s_r&=&-k_B\int d{\bf r}\ p_r({\bf r})\ln (V^n p_r({\bf r})) \label{eq:s_r_app}, \\ 
s_R&=&-k_B\int d{\bf R}\ p_R({\bf R})\ln (V^N p_R({\bf R}))\label{eq:s_R_app}.
\end{eqnarray} 
Indeed, by introducing Eq.~\ref{eq:omega_app} in Eq.~\ref{eq:s_R_app} $s_R$ can be rewritten as 
\begin{eqnarray}
\label{eq:s_R_app_rew}
s_R &=& -k_B\int d{\bf R}\ \left[\int d{\bf r}\ p_r({\bf r})\delta({\bf M}({\bf r}) - {\bf R}))\right]\ln (V^N p_R({\bf R})) \nonumber \\
&=&-k_B\int d{\bf r}\ p_r({\bf r})\ln (V^N p_R({\bf M}({\bf r}))).
\end{eqnarray}
Subtracting Eq.~\ref{eq:s_r_app} and~\ref{eq:s_R_app_rew} results in
\begin{eqnarray}
s_R-s_r=k_B\int d{\bf r}\ p_r({\bf r})\ln \left(\frac{V^{n-N}p_r({\bf r})}{p_R({\bf M}({\bf r}))}\right),
\end{eqnarray}

and by virtue of Eq.~\ref{eq:pbar_app} and~\ref{eq:omega1dec}, one finally obtains
\begin{equation}
s_R-s_r=S_{map},
\end{equation}
further highlighting that the mapping entropy represents the difference in information content between the distribution obtained
by reducing the level of resolution at which the system is observed, $p_R({\bf R})$, and the original, microscopic reference, $p_r({\bf r})$.

\section{Explicit calculation of the mapping entropy}
\label{app:derivation}

We here provide full detail of our derivation of the mapping entropy, as in Eqs.~\ref{eq:smapint}-\ref{eq:smap_reweighted}, and its cumulant expansion approximation, Eq.~\ref{eq:smap_cum.1_main}, starting from Eq.~\ref{eq:smap_app}.

In the case of CG representations obtained by decimating the number of original degrees of freedom of the system, the mapping entropy $S_{map}$ in Eq.~\ref{eq:smap_app} vanishes if the probabilities of the microscopic configurations that map onto the same CG one are the same \cite{rudzinski_2011,foley2015impact}. In the canonical ensemble, the requirement is that those configurations must possess the same energy. This can be directly inferred by writing the negative of the average in Eq~\ref{eq:smap_app} as
\begin{eqnarray} 
\label{eq:ururprime}
&&\left\langle \ln \left[ \frac{\bar{p}_r({\bf r})}{p_r({\bf r})} \right] \right\rangle = \int d{\bf r}\ p_r({\bf r}) \times\\ \nonumber
&&\ln\left[\frac{\int d{\bf r}'\exp[-\beta(u({\bf r}')-u({\bf r}))]\delta({\bf M}({\bf r}') - {\bf M}({\bf r}))}{\int d{\bf r}'\delta({\bf M}({\bf r}') - {\bf M}({\bf r}))}\right],
\end{eqnarray}
so that if $u({\bf r}')=u({\bf r})~\forall~{\bf r}'~\text{s.t.}~{\bf M}({\bf r}')={\bf M}({\bf r})$, the argument of the logarithm is unity and the right-hand side of Eq.~\ref{eq:ururprime} vanishes.

Importantly, this implies that no information on the system is lost along the coarse-graining procedure if CG macrostates are generated by grouping together microscopic configurations characterized by having the same energy. In our case, this translates into the search for \emph{isoenergetic mappings}.

By introducing $1=\int d{\bf R}\ \delta({\bf M}({\bf r}) - {\bf R})$ in Eq.~\ref{eq:ururprime}, one obtains
\begin{eqnarray}
&&S_{map} = -k_B \int d{\bf R}\ \int d{\bf r}\ p_r({\bf r}) \delta({\bf M}({\bf r}) - {\bf R}) \times \\ 
&&\ln\left[\frac{\int d{\bf r}'\exp[-\beta(u({\bf r}')-u({\bf r}))]\delta({\bf M}({\bf r}') - {\bf R})}{\int d{\bf r}'\delta({\bf M}({\bf r}') - {\bf R})}\right] \nonumber \\ 
\label{eq:mapentr}
&&=\int d{\bf R}\ p_R({\bf R})S_{map}({\bf R}),
\end{eqnarray}
so that the overall mapping entropy is decomposed as a weighted average over the CG configuration space of the mapping entropy $S_{map}({\bf R})$ of a \emph{single}
CG macrostate,
\begin{eqnarray}
\label{eq:smapr}
&&S_{map}({\bf R})=-\frac{k_B}{\ p_R({\bf R})}\int d{\bf r}\ p_r({\bf r}) \delta({\bf M}({\bf r}) - {\bf R}) \times \\ \nonumber
&&\ln\left[\frac{\int d{\bf r}'\exp[-\beta(u({\bf r}')-u({\bf r}))]\delta({\bf M}({\bf r}') - {\bf R})}{\int d{\bf r}'\delta({\bf M}({\bf r}') - {\bf R})}\right].
\end{eqnarray}

Eq.~\ref{eq:smapr} shows that determining $S_{map}({\bf R})$ for a given macrostate ${\bf R}$ involves a comparison of the energies of all pairs of microscopic configurations that map onto it. A further identity $1=\int d U'\delta(u({\bf r}')-U')$ fixing the energy of configuration ${\bf r'}$ can be inserted in the logarithm of Eq.~\ref{eq:smapr} to switch from a configurational to an energetic integral. This provides:
\begin{eqnarray}
\label{eq:firstdeltaU}
&&\ln\left[\frac{\int d{\bf r}'\exp[-\beta(u({\bf r}')-u({\bf r}))]\delta({\bf M}({\bf r}') - {\bf R})}{\int d{\bf r}'\delta({\bf M}({\bf r}') - {\bf R})}\right]= \nonumber \\ 
&&\ln\int d U'P(U'|{\bf R})\exp[-\beta(U'-u({\bf r}))],
\end{eqnarray}
where
\begin{equation}
\label{eq:app_microcanonic}
P(U'|{\bf R}) = \frac{\int d{\bf r}'\delta({\bf M}({\bf r}') - {\bf R})\delta(u({\bf r}')-U')}{\int d{\bf r}'\delta({\bf M}({\bf r}') - {\bf R})}
\end{equation}
is the microcanonical (unweighted) conditional probability of possessing energy $U'$ given that the CG macrostate is ${\bf R}$. It is possible to write it as ${\Omega_1(U',{\bf R})}/{\Omega_1({\bf R})}$, that is, the multiplicity of AA configurations such that ${\bf M}({\bf r}) = {\bf R}$ and $u({\bf r}') = U'$ normalized by the multiplicity of configurations that map to ${\bf R}$.

A second identity $1=\int d U\delta(u({\bf r})-U)$ on the energies provides the following expression for $S_{map}({\bf R})$:
\begin{eqnarray}
\label{smapR_second_delta}
&&S_{map}({\bf R})=-k_B\int d{\bf r}\frac{p_r({\bf r})}{p_R({\bf R})} \delta({\bf M}({\bf r}) - {\bf R}) \times \nonumber \\ 
&&\ln \left[ \int d U'P(U'|{\bf R})\exp[-\beta(U'-u({\bf r}))] \right] \nonumber \\
&&=-k_B\int dU\ln\left[\int d U'P(U'|{\bf R})\exp[-\beta(U'-U)]\right] \times \nonumber \\
&&\int d{\bf r}\frac{p_r({\bf r})}{p_R({\bf R})}\delta({\bf M}({\bf r}) - {\bf R})\delta(u({\bf r})-U).
\end{eqnarray}
The last integral in Eq \ref{smapR_second_delta}, which we dub $P_{\beta}(U|{\bf R})$,
\begin{equation}
\label{eq:app_canonic}
P_{\beta}(U|{\bf R})=\int d{\bf r}\frac{p_r({\bf r})}{p_R({\bf R})}\delta({\bf M}({\bf r}) - {\bf R})\delta(u({\bf r})-U)
\end{equation}
is now the \text{canonical}---i.e., Boltzmann-weighted---conditional probability of possessing energy $U$ provided that ${\bf M}({\bf r}) = {\bf R}$, namely ${p_R(U,{\bf R})}/{p_R({\bf R})}$. One thus obtains:
\begin{eqnarray}
\label{eq:smap_pre_rew}
&&S_{map}({\bf R})=-k_B\int dU P_{\beta}(U|{\bf R})\times  \\
&&\ln\left[\int d U'P(U'|{\bf R})\exp[-\beta(U'-U)]\right]\nonumber 	\\
&&= -k_B\ln\left[\int d U'P(U'|{\bf R})\exp[-\beta(U'-\langle U\rangle_{\beta|{\bf R}})]\right],\nonumber
\end{eqnarray}
where
\begin{equation}
\langle U\rangle_{\beta|{\bf R}}=\int dU P_{\beta}(U|{\bf R})U
\end{equation}
is the canonical average of the microscopic potential energy over the CG macrostate ${\bf R}$.

A direct calculation of $S_{map}({\bf R})$ starting from the last line of Eq.~\ref{eq:smap_pre_rew} requires to perform an average over the microcanonical distribution $P(U'|{\bf R})$, which is not straightforwardly accessible in NVT simulations. However, there is a connection between $P(U|{\bf R})$ in Eq.~\ref{eq:app_microcanonic} and $P_{\beta}(U|{\bf R})$ in Eq.~\ref{eq:app_canonic}: if one writes ${p_R({\bf R})}$ as $\int d U' \exp[-\beta(U')] \Omega_1(U', {\bf R})$ and $p_R(U, {\bf R})$ as $\exp[-\beta(U)] \Omega_1(U, {\bf R})$, standard reweighing provides
\begin{equation}
\label{eq:can_to_mc}
P(U|{\bf R})=\frac{P_{\beta}(U|{\bf R})\exp[\beta U]}{\int dU'P_{\beta}(U'|{\bf R})\exp[\beta U']}.
\end{equation}
Eq. \ref{eq:can_to_mc} enables one to convert the microcanonical average in Eq.~\ref{eq:smap_pre_rew} to a canonical one, so that
\begin{equation}
\label{eq:smap_reweighted_app}
\hspace*{-0.1cm}
S_{map}({\bf R}) = k_B \ln\left[\int dU' P_{\beta}(U'|{\bf R}) \ e^{\beta(U' - \langle U\rangle_{\beta|{\bf R}})} \right].
\end{equation}
Finally, by means of a second order cumulant expansion of Eq. \ref{eq:smap_reweighted} one obtains
\begin{equation}
S_{map}({\bf R})\simeq k_B\frac{\beta^2}{2}\langle(U-\langle U\rangle_{\beta|{\bf R}})^2\rangle_{\beta|{\bf R}},
\end{equation}
that inserted in Eq.~\ref{eq:mapentr} results in a \emph{total} mapping entropy given by Eq. \ref{eq:smap_cum.1_main}.

\bibliographystyle{ieeetr}
\bibliography{main.bib}

\end{document}